\documentclass[twocolumn]{aas63_mod}
\usepackage[caption=false]{subfig}
\usepackage{booktabs}
\usepackage[T1]{fontenc}
\usepackage{lineno}
\usepackage{longtable}
\usepackage{afterpage}
\usepackage{listings}
\definecolor{rotundaorange}{rgb}{0.898,0.44705882352,0}
\definecolor{pantherpurple}{rgb}{0.2,0,0.447058}
\definecolor{raiderred}{rgb}{0.796875,0,0}
\definecolor{remeisblue}{rgb}{0, 0, 0.5450980}

\newcommand{\msun}{$M_{\odot}$}
\newcommand{\msundyn}{\ifmmode M_{\odot}\else$M_{\odot}$\fi}
\newcommand{\rsun}{\ifmmode R_{\odot}\else$R_{\odot}$\fi}
\newcommand{\degree}{\ifmmode^{\circ}\else$^{\circ}$\fi}
\newcommand{\amin}{\ifmmode^{\prime}\else$^{\prime}$\fi}
\newcommand{\asec}{\ifmmode^{\prime\prime}\else$^{\prime\prime}$\fi}

\hypersetup{linkcolor=xlinkcolor,citecolor=xlinkcolor}

\newcommand{\tzx}{$T_{0, x}$}
\newcommand{\st}{\texttt{SPIDER\_TWISTER}}
\newcommand{\pint}{\texttt{PINT}}
\newcommand{\gap}{gap limit}
\newcommand{\xro}{\texttt{XR1}}
\newcommand{\xrt}{\texttt{XR2}}
\newcommand{\tox}{\texttt{T0X}}
\newcommand{\aox}{\texttt{A1X}}
\newcommand{\dt}{$\Delta T_{0}$}
\newcommand{\pk}{\texttt{pintk}}
\newcommand{\ti}{$t_i$}
\newcommand{\trd}{$t_{\Delta_{R},i}$}
\newcommand{\pb}{$P_{\rm b}$}
\newcommand{\pbdot}{$\dot{P_{\rm b}}$}
\newcommand{\ra}{$\alpha$}
\newcommand{\dec}{$\delta$}

\begin{document}

\title{A Novel Technique for Long-term Timing of Redback Millisecond Pulsars}

\correspondingauthor{Kyle A. Corcoran}
\email{kcorcoran@cpi.com}

\author[0000-0002-2764-7248]{Kyle A. Corcoran}
\altaffiliation{Computational Physics, Inc.}
\affiliation{University of Virginia, Department of Astronomy, Charlottesville, VA 22904, USA}
\author[0000-0001-5799-9714]{Scott M. Ransom}
\affiliation{National Radio Astronomy Observatory, Charlottesville, VA 22903, USA}
\author[0009-0001-2223-2975]{Alexandra C. Rosenthal}
\affiliation{University of Virginia, Department of Astronomy, Charlottesville, VA 22904, USA}

\author[0000-0002-2185-1790]{Megan E. DeCesar}
\affiliation{George Mason University, Fairfax, VA 22030, USA}
\affiliation{Resident at the U.S. Naval Research Laboratory, Washington, DC 20375, USA}

\author[0000-0003-1307-9435]{Paulo C. C. Freire}
\affiliation{Max-Planck-Institut für Radioastronomie, Auf dem Hügel 69, D-53121 Bonn, Germany}

\author[0000-0003-2317-1446]{Jason W. T. Hessels}
\affiliation{Anton Pannekoek Institute for Astronomy, University of Amsterdam, Science Park 904, 1098 XH Amsterdam, The Netherlands}
\affiliation{ASTRON, Netherlands Institute for Radio Astronomy, Oude Hoogeveensedijk 4, 7991 PD Dwingeloo, The Netherlands}

\author[0000-0001-5229-7430]{Ryan S. Lynch}
\affiliation{Green Bank Observatory, P.O. Box 2, Green Bank, WV 24494, USA}

\author[0000-0001-5624-4635]{Prajwal V. Padmanabh}
\affiliation{Max Planck Institute for Gravitational Physics (Albert Einstein Institute), D-30167 Hannover, Germany}
\affiliation{Leibniz Universität Hannover, D-30167 Hannover, Germany}

\author[0000-0001-9784-8670]{Ingrid H. Stairs}
\affiliation{Department of Physics and Astronomy, University of British Columbia, 6224 Agricultural Road, Vancouver, BC V6T 1Z1 Canada}





\begin{abstract}
We present timing solutions spanning nearly two decades for five redback (RB) systems found in globular clusters (GC), created using a novel technique that effectively ``isolates'' the pulsar.  By accurately measuring the time of passage through periastron ($T_0$) at points over the timing baseline, we use a piecewise-continuous, binary model to get local solutions of the orbital variations that we pair with long-term orbital information to remove the orbital timing delays.  The isolated pulse times of arrival can then be fit to describe the spin behavior of the millisecond pulsar (MSP).  The results of our timing analyses via this method are consistent with those of conventional timing methods for binaries in GCs as demonstrated by analyses of NGC 6440D.  We also investigate the observed orbital phase variations for these systems.  Quasi-periodic oscillations in Terzan 5P's orbit may be the result of changes to the gravitational-quadruple moment of the companion as prescribed by the Applegate model.  We find a striking correlation between the standard deviation of the phase variations as a fraction of a system's orbit ($\sigma_{\Delta T_0}$) and the MSP's spin frequency, as well as a potential correlation between $\sigma_{\Delta T_0}$ and the binary's projected semi-major axis.  While long-term RB timing is fraught with large systematics, our work provides a needed alternative for studying systems with significant orbital variations, especially when high-cadence monitoring observations are unavailable.
\end{abstract}




\section{\bf{Introduction}} \label{sec:intro}
Millisecond pulsars (MSPs) with binary companions are an interesting class of radio pulsars and unique laboratories for various tests of fundamental physics.  These neutron stars have been spun up via accretion of material from a close binary companion \citep{Alpar1982}, and we most commonly observe the resultant system in a relatively wide orbit with a low-mass, white dwarf companion (\citealt{Manchester2005,Tauris2006}, for a survey of this topic see \citealt{Tauris2023}). 

\subsection{``Redback'' pulsars}

A portion of systems with other companion types can evolve to much more compact orbits ($P_{\rm b}<1\,{\rm day}$), and one such sub-class is called Redbacks (RBs).  These MSPs are members of the ``spider'' pulsar family and have H-rich, non-degenerate companions that have masses typically between $0.1M_{\odot}<M_{c}<0.9M_{\odot}$ \citep{Roberts2013,Strader2019}.  RBs typically have orbital periods of only a few hours, and their neutron star masses are generally found to exceed $1.4\,M_{\odot}$ \citep{Strader2019}. Additionally, RBs are exceptional producers of multi-wavelength emission, producing features such as radio pulsations, $\gamma$-ray pulsations \citep{Deneva2021,Thongmeearkom2024}, and optical light-curve variations \citep{Bellm2016,Yap2023}.

Due to their compact nature, binary interactions play a large role in the evolution and observed properties of RBs.  The intense pulsar wind ablates material from the companion, creating circumbinary, ionized material.  This material can cause radio eclipses -- most likely  caused by synchrotron absorption \citep{Polzin2018} -- that can shroud radio pulsations for significant portions of the orbit \citep{Nice1990} around superior conjunction.  These eclipses can be highly irregular, though, and dependencies on observing frequency can further impact the duration and appearance of any one eclipse \citep[e.g.,][]{Nice1990,You2018}.  Additional eclipse-like events due to material at other orbital phases can also mask pulsations \citep[e.g.,][]{Bilous2019}, and the long-term behavior of the binary orbit in RBs has been observed to wander drastically \citep{Prager2017PhDT,Clark2021,Thongmeearkom2024}.  These factors limit the accuracy of binary models to describe the system over time.

Timing studies of MSPs in binaries given sufficient baselines can yield insights into interesting physics in areas such as general relativity
(e.g., \citealt{Kramer2021}, see \citealt{Freire_Wex_2024} for a review) and the neutron star equation of state
\citep{Fonseca2021}; however, the barrage of impediments listed above means that despite their potentially large NS masses, RBs are often excluded from consideration for long-term timing efforts in the radio.  Obtaining successive detections -- let alone detections over short to long baselines -- of the MSP can be difficult, and linking observations together in a way that yields insights into the physical characteristics of these systems is also non-trivial.  While the binary interactions in RBs may not make them ideal test beds for probing the most impactful binary physics, accurate measurements of physical properties derived from long-baseline timing still would yield important limits for modeling the companion's interior and binary evolution modeling.  These limits can inform a variety of interesting cases.  The majority of published radio-timing solutions for RBs, however, only cover a baseline of a few years 
\citep{Archibald2013,Prager2017PhDT,MiravalZanon2018,Deneva2021,Padmanabh2024,Ghosh2024}. \citet{Thongmeearkom2024} achieved phase connection using $\gamma$-ray data from the \textit{Fermi} Large Area Telescope (LAT), which spans 15 years, but this technique requires that the pulsars are bright in $\gamma$ rays. \citet{Nice2000} and \citet{Ridolfi2016} presented radio-timing solutions for around a decade of targeted observations of two redback systems in the globular clusters Terzan 5 and 47 Tucanae, B1744$-$24A (henceforth Ter5A) and PSR~J0024$-$7204W (47~Tuc~W). Targeted radio observations or even archival radio data spanning this duration or longer for individual RBs are generally not available, making these long-term studies difficult to conduct.

 However, the main obstacle preventing long-term timing studies of RBs is the limitations of the  orbital timing models. The BTX model, which is the one currently used to describe the orbital evolution of RBs, uses multiple orbital frequency derivatives in a Taylor expansion of the orbital phase with time.  At higher orders, of which RBs have many, BTX becomes numerically unstable and yields non-physical predictions for the variation of the orbital period.  This also makes datasets with large gaps and/or long-term observations very difficult to fit. Models which successfully fit long time spans often completely fail to describe and phase connect new observations, requiring specialized techniques to recover phase connection.

\begin{figure*}
    \centering
    \includegraphics[width=\textwidth]{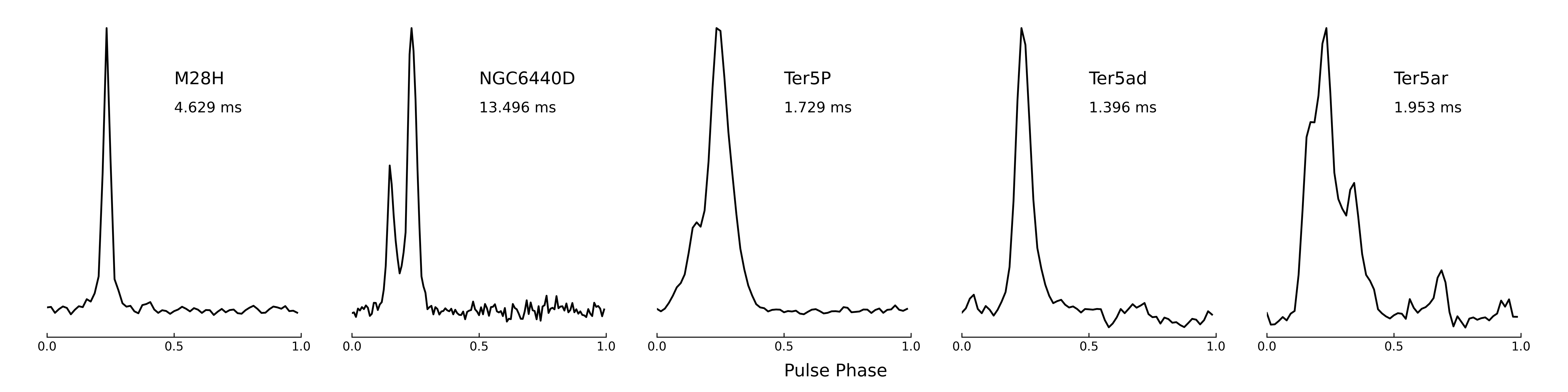}    
    \caption{Summed pulse profiles from coherently dedispered 2\,GHz observations for each of the five RBs analyzed in this work.}
    \label{fig:pulse_profiles}
\end{figure*}

\subsection{Globular cluster pulsars}

Globular clusters (GCs) present an opportunity to provide long-term baselines for RBs by exploiting observations of the cluster aimed at the pulsar population.
Per unit of stellar mass, GCs have more than 3 orders of magnitude more X-ray binaries and MSPs than the Galactic disk; this suggests that these systems formed dynamically, which is possible in GCs because of their very large stellar densities \citep{Ransom2008,Freire2013}.
One of the interesting features of the pulsar population in GCs is that, in the densest of them, exchange encounters can happen after the pulsar was recycled, allowing the formation of systems that are unlike any formed in
the Galactic disk. 
This applies to RBs as well: Unlike in the case of the Galactic RBs, some of the RBs in GCs might be
the product of pair exchange \citep[where the star that spun up the pulsar is ejected and replaced with a more massive star;][]{Ransom2005,Prager2017PhDT}. In any case, GCs have a disproportionately large number of RBs compared to the Galactic pulsar population.

Given the very large distances of GCs, only a small fraction of their MSPs has been discovered, with the majority remaining undetected due to sensitivity limitations. This means that observations with improved sensitivity allow the discovery of
new pulsars;
they also result in improved timing of previously known pulsars.
One of the benefits of timing pulsars in GCs is to use pulsars as accelerometers, which allow constraints on GC mass models. This is especially true of GCs that contain a large population of pulsars, like Terzan 5 \citep{Prager2017} and 47~Tucanae \citep{Freire2017,Abbate2018}. 

Terzan 5 (hereafter Ter5) has the largest known population of pulsars in any GC \citep[49;][]{Padmanabh2024} with over half being in binaries -- four of which are RBs. Ter5A is the first ever RB discovered \citep{Lyne1990,Nice1990,Nice2000} and one of the most compact RBs known; Ter5ad is the record holder for the fastest spinning MSP  \citep[$P\sim1.39\,\rm{ms}$;][]{Hessels2006}; Ter5P is also rapidly spinning ($P\sim1.72\,\rm{ms}$), has the second largest mass fraction for RBs due to its more massive companion ($M_{\rm{c, min}}\sim0.38\,M_{\odot}$) and Ter5ar is the most recently discovered, rapidly rotating ($P\sim1.95\,\rm{ms}$) RB with a larger mass companion ($M_{\rm{c, min}}\sim0.34\,M_{\odot}$) as well \citep{Padmanabh2024}.  A smaller population of binary pulsars is found in M28, and it, too, contains RBs -- M28I, which is a transitional MSP \citep[switches between a radio pulsar state and a low-mass X-ray binary state;][]{Papitto2013}, and M28H, which is another RB thought to be the product of pair exchange \citep{Pallanca2010,Bogdanov2011}.  NGC 6440, which has the smallest binary population of the three clusters mentioned here, also hosts a RB, NGC 6440D \citep{Freire2008}, that exhibits relatively stable orbital variations  (Ransom et al., in prep).  These three GCs have been actively monitored for nearly the past 20 years through observations at various frequencies using the Green Bank Telescope (GBT), making them unique sources to test RB timing methods on a longer scale than has previously been possible.

\subsection{Structure of this work}

In \textsection\ref{sec:obs} we briefly describe the roughly 20 years of archival data from the GBT.  In \textsection\ref{sec:methods} we present a novel technique for ``isolating'' the MSP from the orbital effects of the binary companion, allowing us to time the underlying, highly accurate clock.   This method provides the advantage of phase connecting long-term observational datasets even with large gaps since the timing of an ``isolated'' pulsar is predictive, effectively overcoming the aforementioned limitations of the BTX model.
In \textsection\ref{sec:results} we present our long-term timing analysis of five RBs in GCs -- Ter5P, Ter5ad, Ter5ar, M28H, and NGC 6440D, our fully phase-connected timing solutions, the long-term orbital variations that were removed to create them, and pulse profiles (shown in Figure \ref{fig:pulse_profiles}) for each system.
Apart from NGC 6440D, none of these pulsars had any published timing solutions until now. A separate publication
on 34 years of timing of another RB, Ter5A, that uses the same technique is presented in \citet{Rosenthal2024}.   We then discuss the results from our technique and how they compare to those of conventional timing methods, the quasi-periodic oscillations seen in Ter5P, and a correlation between orbital variations and spin frequency in \textsection\ref{sec:discussion}.  Finally, we summarize our work in \textsection\ref{sec:summary}.

\section{\bf{Observations}} \label{sec:obs}
We used archival data for each cluster from observations with the GBT spanning roughly 20 years.  A majority of our data were taken using S-band (1600-2400\,MHz) and L-band (1100-1900\,MHz) receivers with $\sim$600-700\,MHz of useable bandwidth in each case.  Additionally, a smaller fraction of data were taken at 820\,MHz.  In our data, the GBT Pulsar SPIGOT \citep{SPIGOT} backend was used for all observations prior to MJD 55000, the GUPPI \citep{GUPPI} backend was then used for observations up until MJD 58933, and that same day the VEGAS \citep{VEGAS} backend was switched onto the GBT.  Consequently, we acquired observations with both GUPPI and VEGAS on the day of the switchover.  The SPIGOT data were taken using incoherent dedispersion, and detailed information about the observations for Ter5, which were obtained in a consistent manner to the observations for M28H and NGC 6440, can be found in \citet{Ransom2005}.  Prior to MJD 55422, observations using GUPPI (only a few scans for each system) were also obtained with incoherent dedispersion; the remaining GUPPI and VEGAS observations were taken using coherent dedispersion, and detailed information about the observations for Ter5 can be found in \citet{Martsen2022}.  As with SPIGOT, the GUPPI and VEGAS observations were obtained in a consistent manner to those for M28H and NGC 6440, with the obvious exception of the dispersion measure (DM) where the coherent data are dedispersed.

\section{\bf{Timing Methodology}} \label{sec:methods}
In the text that follows we introduce the prescription for and our workflow implementation of the Ransom-O'Neill Isolation (ROI)\footnote{Named by K.A.C. as the upfront work to determine accurate \tzx\ measurements yields a significant return on investment.} technique.  To briefly outline the process: we first measure the change in the time of periastron passage as a function of time (see, \textsection\ref{subsec:detections}-\ref{subsec:btx-gp}), which can be done through various methods \citep[e.g.,][]{Shaifullah2016,Voisin2020,O'Neill2025,Rosenthal2024}.  Then, we properly remove the Roemer delay from the pulse times of arrival (TOAs) for each observation/group of observations (see, \textsection\ref{subsec:roi} \& Appendix \ref{appendix:code}).  Finally, we time the rotation of the pulsar as if it were isolated (see \textsection\ref{subsec:fitting}).  While we use this method here to time RBs found in GCs, we note that the technique is also applicable to Galactic spider systems, and could theoretically be used for ``regular'' pulsar binaries (although since the method doesn't jointly fit for orbital parameters when fitting for other timing parameters, it would be sub-optimal for such work).

\subsection{Detections and Initial \tzx\ Measurements}\label{subsec:detections}
We first needed to identify all observations for which we had detections of the pulses for each system in our observations.  For RBs, this process can be difficult as the $T_0$ value -- the time of passage through periastron, which for our circular ($e=0$) models is when the MSP crosses the plane of the sky moving away from the observer, called the time of the ascending node -- of successive detections can vary dramatically over time (see Figure \ref{fig:deltaT0}).  We used \st\footnote{\url{https://github.com/alex88ridolfi/SPIDER_TWISTER}} \citep{Ridolfi2016}, which performs searches in orbital phase to return the most probable $T_0$ value for a given observation, to obtain detections for each source.  As our data span roughly 20 years, we automated our searches by allowing \st\ to search $\pm10\%$ in orbital phase in the time-series for each observation from its associated $T_0$ predicted by an initial set of parameters for each system.  Even for systems with significant orbital wander, we found this search setup to be effective in finding all available detections in its dataset.  We inspected the output plots of each time-series folded at the returned $T_0$ value\footnote{We note that \tzx\ here refers to $T_0$ for each observation.  Herein we generally use this to refer to the $T_0$ value associated with an arbitrary, binary-piecewise group, $x$.} (\tzx) and noted non-detections to discard for subsequent steps.  At this stage, the \tzx\ measurements are not necessarily the most accurate value for achieving ROI; however, these measurements are precise enough for obtaining TOAs from each observation.  These measurements thus served as an initial set of values that we refined throughout the process (see panel (a) in Figure \ref{fig:T0_progression}).  
\vfill

\subsection{Producing TOAs} \label{subsec:toas}
We folded the time-series data, using the \texttt{prepfold} routine from PRESTO\footnote{\url{https://github.com/scottransom/presto}} \citep{ransom2001,presto}, for each detection of each MSP using its predicted spin period, DM value, long-term average orbital parameters, and the \tzx\ measurement for each observation obtained from \st.  For Ter5P, Ter5ad, Ter5ar, and M28H, we then integrated over set intervals (10\,min for Ter5P and M28H \& 30\,min for Ter5ad and Ter5ar) to obtain TOAs for each system.  Due to the extremely rapid spin periods of Ter5ad and Ter5P, we used separate pulse templates for data obtained in incoherent and coherent dedispersion modes, allowing us to mitigate the effects of smearing for these systems.  We visually inspected each fold and noted areas to avoid where there was no pulse (e.g., from both regular and irregular eclipses) as well as areas that may produce erroneous TOAs (e.g., when strong interference was present).  We then used PRESTO's \texttt{get\_TOAs.py} routine to extract TOAs for each observation in the areas where a signal was present.  Finally, we discarded TOAs with errors larger than $30\,\mu\rm{s}$.

In the case of NGC 6440D, we used the TOAs from Ransom et al. (in prep) with errors less than $30\,\mu\rm{s}$.  These are produced by determining the number of TOAs that could be obtained with sufficient S/N in each observation and integrating over intervals that yield this number of TOAs (roughly 4-15\,min integrations).  We chose to maintain this slight difference in methodology rather than re-integrating at a set interval to keep the TOAs consistent between both analyses, allowing us to directly compare results obtained via traditional timing techniques to those obtained using the ROI technique.

\subsection{Updating Orbital Properties}\label{subsec:orbits}
To ensure that our long-term description of the binary orbital properties was accurate, we used the \tzx\ values from \st\ to determine the average orbital period of each system.  For all systems except Ter5P and NGC 6440D, we assumed that the period of a circular orbit was constant over the baseline of our observations.  We also assumed for all systems that the semi-major axis was constant (see Appendix \ref{sec:appendixdynamics} for more discussion on these assumptions).  We then computed the difference between our measured \tzx\ value and the value that would be predicted by a constant orbital period and its associated reference epoch $T_{0, \rm{ref}}$.  With a constant orbital period, these \dt\ values over time will show long-term, linear trends if an adjustment to \pb\ was necessary.  We fit a linear trend in these cases as:
\begin{equation}
    \frac{\Delta T_0}{t P_b} = \frac{\Delta T_0 f_{b}}{t}= \frac{\Delta\phi_{b}}{t},
\end{equation}
which describes the change in phase ($T_0$ in this case) over time, $t$, which we used to update the period by computing the change to the orbital frequency, $f_b$, via
\begin{equation}
    f_{b, \rm{new}} = f_{b, \rm{old}} + \frac{\Delta T_0}{t} f_{b, \rm{old}} 
\end{equation}
and then using
\begin{equation}
    P_{b, \rm{new}} = \frac{1}{f_{b, \rm{new}}}.
\end{equation}
In the cases of Ter5P and NGC 6440D, the dominant \dt\ trend was a quadratic; therefore, we fit both a quadratic and linear term to correct the \pb\ measurement.

\subsection{Constructing Piecewise-Continuous Groups \& Parameter Files} \label{subsec:initpars}
For RBs, it is sometimes possible to use a piecewise-discontinuous model to time the system over long baselines, wherein individual chunks of overlapping TOAs spanning some time frame can be strung together with fits using independent models with constant orbital parameters \citep[e.g.,][]{Blandford1976} to achieve a connected, long-term timing solution \citep[see][]{Archibald2013,Shaifullah2016,Rosenthal2024}.  A piecewise-continuous, binary model, though, allows for computation of timing residuals for all TOAs over the duration of the baseline with a proper and changing orbit without need for independent fits of other parameters \citep[e.g.,][]{Wex1998,Wex1998b,O'Neill2025}.  This process is particularly useful in cases where orbital variations make connecting overlapping chunks difficult.  As such, we opted to use the \texttt{BT\_piecewise} model \citep[see][for full details on this model]{O'Neill2025} inside the pulsar timing package, \pint\footnote{\url{https://github.com/nanograv/PINT}}\ \citep{PINT} in isolating the TOAs for each system.  While this is not necessary for removing the timing delays described in \textsection\ref{subsec:roi}, it is useful in predicting how the isolated timing residuals will look with those delays removed.

Using the TOAs from \textsection\ref{subsec:toas}, we constructed an initial set of piecewise-continuous groups of similar TOAs for each system.   We used the \texttt{TOAs.get\_clusters} method inside \pint, to identify groups of temporally related TOAs over a specified \gap.  As our datasets contain varying sizes of gaps between observations, we tested a number of different \gap\ values using TOAs from Ter5P and Ter5ar by iteratively refining the piecewise groups and \tzx\ measurements.  Ultimately, we found that a \gap\ of 0.5\,d worked well for most systems to group observations temporally very close together (e.g., two scans on the same day) while not creating groups too temporally large to get an accurate picture of the local solution.  Due to the nature of how NGC 6440D's TOAs were constructed, we used a \gap\ of only 0.04166\,d to create its piecewise groups as larger \gap\ values created some groups with TOAs too sparse to describe the local orbital variations.  Once all of the TOAs were assigned a group number, we generated four values for each group that the binary-piecewise model uses to describe local solutions: \xro, \xrt, \aox, and \tox, which are the start, finish, $a\sin\left({i}\right)/c$, and \tzx\ values, respectively.  It is important to note again that we assumed the semi-major axis of the binary is constant over our baseline for these systems.


\begin{figure*}
    \centering
    \includegraphics[width=\linewidth]{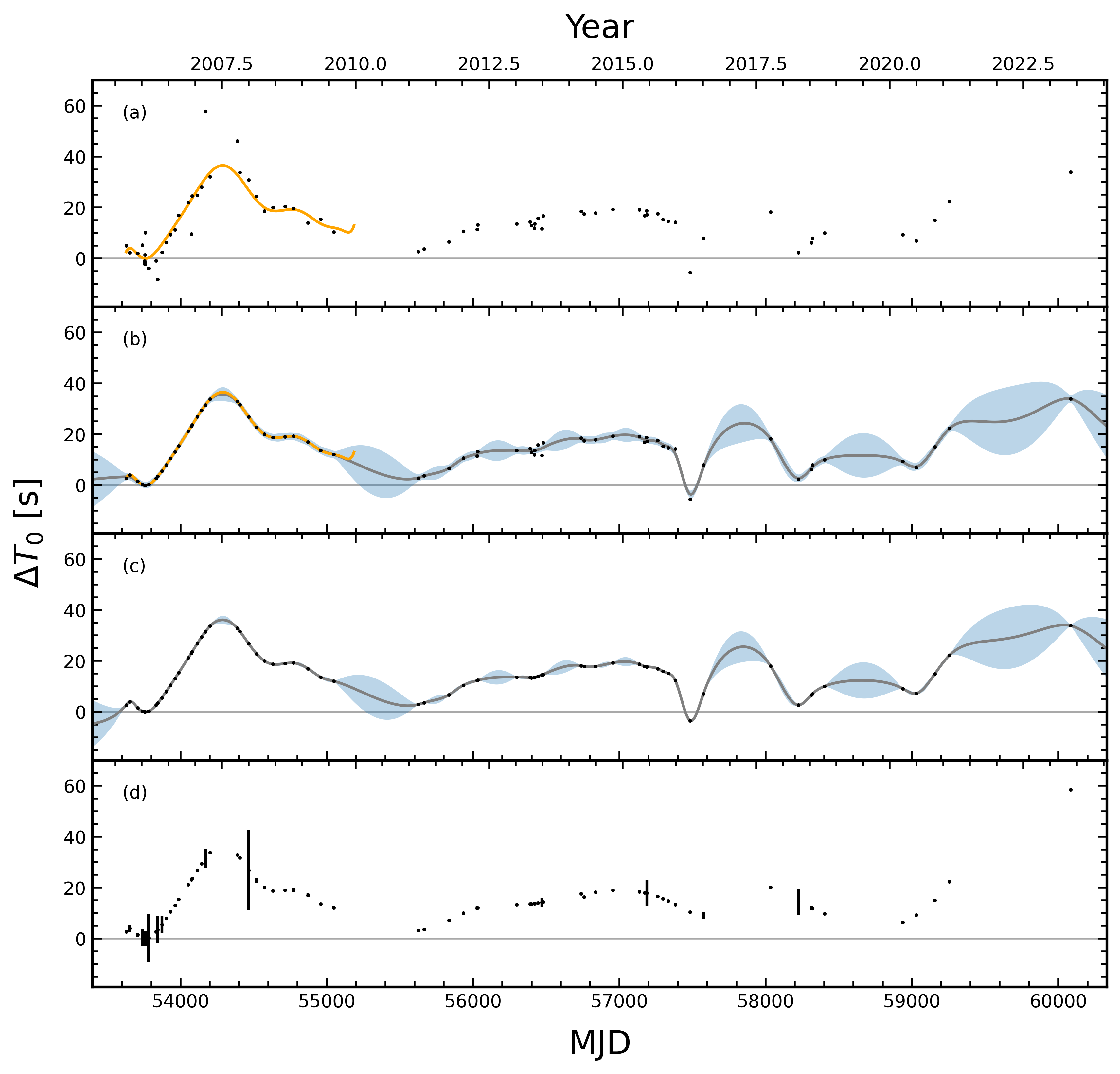}
    \caption{Each panel shows different measurements of time of periastron passage ($T_0$) deviations over time for M28H as black points.  In panel (a), we show the \dt\ values derived from the \st\ values described in \textsection\ref{subsec:detections}, and we overlay in orange the predicted values from the phase-connected, BTX model described in \textsection\ref{subsec:btx-gp}.  In panel (b), we show the \dt\ values accounting for BTX predictions, as well as a Gaussian process regression (GPR) to interpolate between the BTX-informed, $T_0$ values and the remaining \st\ measurements.  In panel (c), we show the remaining values set to those GPR predictions, as well as a new regression to describe the updated measurements.  This process still benefits from manual improvement to ensure \dt\ is relatively smooth with time, and we show the resultant values and their errors in panel (d).}
    \label{fig:T0_progression}
\end{figure*}

We then created three different parameter files (or par files) that describe each system -- one that describes the local, binary solution of the piecewise groups, another the long-term, binary solution, and finally the isolated, spin behavior.  To successfully track the rotations of the pulsar (phase connect) over the full timing baseline, it is helpful (although not strictly necessary) that these three par files be derived from a solution that has some portion of the data phase connected.  We initially started with right ascension, declination, the spin frequency and its first derivative from previous short-duration phase connection from \citet{Ransom2005} \& \citet{Padmanabh2024} for Ter5 systems, \citet{Begin2006} for M28H, and Ransom et al. (in prep) for NGC 6440D.  We then assumed simple binary models using the \pb, \pbdot (where applicable), and \tzx\ values as determined in \textsection\ref{subsec:orbits} to create the binary-piecewise and long-term-binary (henceforth simple-binary) par files.  For the isolated par file, which only describes the MSP's spin, we simply removed the binary information.


\subsection{Refining \tzx\ with \rm{\texttt{TEMPO}}} \label{subsec:tempo}
As previously stated, the initial measurements of \tzx\ do not always suffice to accurately describe the behavior of each piecewise group.  In general, the processes described in the following section are sufficient for correcting the accuracy of each measurement to proceed with the isolation steps.  For Ter5P, as well as \citeauthor{Rosenthal2024}'s \citeyearpar{Rosenthal2024} work with Ter5A, though, we found that the results of those processes were improved by first refining the \tzx\ measurements that are used as inputs by fitting $T_0$ directly to the measured TOAs.  To do this, we used \texttt{TEMPO}\footnote{\url{https://tempo.sourceforge.net/}}\ and its \texttt{TRACK} capabilities to fit the TOAs for \tzx\ in places where the value measured via \st\ was not sufficient.  In some cases, problematic TOAs that had made it through the initial cleaning stage had caused the initial measurement to be less accurate, while in others, a correction for an improper phase wrap (i.e., pulse count) greatly improved the measurement.  The \tox\ values were then updated in the binary-piecewise par file and used for the final improvement steps.

\subsection{Refining \tzx\ via BTX Information \& Gaussian Process Regression} \label{subsec:btx-gp}
To obtain \tzx\ measurements that were sufficient to describe the local behavior of each group well, we implemented a two-step process to enhance each value's accuracy.  The first was to bring back some of the information that was discarded in creating the simple-binary model.  We used the orbital frequency derivatives from the BTX solutions and phases from \pint's \texttt{model.orbital\_phase} method to compute the orbital period at the current \tzx\ for each group contained in the original phase-connected solution (see panel (a) in Figure \ref{fig:T0_progression}).  For systems where we assumed a constant orbital period, which are all but Ter5P and NGC 6440D, this value did not change.  These values were the driving force behind the earlier solutions for describing the orbit at all points for that data, so naturally they locally described the TOAs in those portions of our data to a high degree of accuracy.  We then updated the \tzx\ and \dt\ values for this subset of the data.

The second step was to use our BTX derived \tzx\ values, the measured \tzx\ values for the remaining groups, and both of their corresponding \dt\ values as inputs to train a Gaussian process regression \citep[GPR;][]{Rasmussen2006}.  We used the Matérn kernel implemented in the \texttt{scikit-learn} package \citep{scikit-learn} with an initial length scale of $\ell=20$, length scale bounds from 1-10$^{4}$\,d, and a white noise kernel with a noise level of 1.5\,s and noise bound from $10^{-2}$-$10^{2}$\,s.  We used the GPR kernel to predict the \tzx\ values for the groups not set by the BTX model (see panel (b) in Figure \ref{fig:T0_progression}), and we used these values to update each \tox\ value for the piecewise groups (see panel (c) in Figure \ref{fig:T0_progression}).  While these values provided great improvement on the accuracy of the \tzx\ measurements, the quality of the initial measurements were not always such that the GPR arrived at a value that properly described the TOAs' local behavior.  In these cases, we visually inspected \dt\ over time overlaid with the results of the GPR. We exploited the fact that the variation in \dt\ over time should be fairly smooth, meaning it should not make large and rapid changes in the orbital phase/period, to determine the adjustments (ranging from only a few fractions of a second up to a few 10s of seconds) to the \tzx\ value needed to correct the local behavior of a group's TOAs (see panel (d) in Figure \ref{fig:T0_progression}).  After iteratively refining the GPR, we updated the \tox\ values for the binary-piecewise model.

\begin{figure*}
    \centering
    \includegraphics[width=\textwidth]{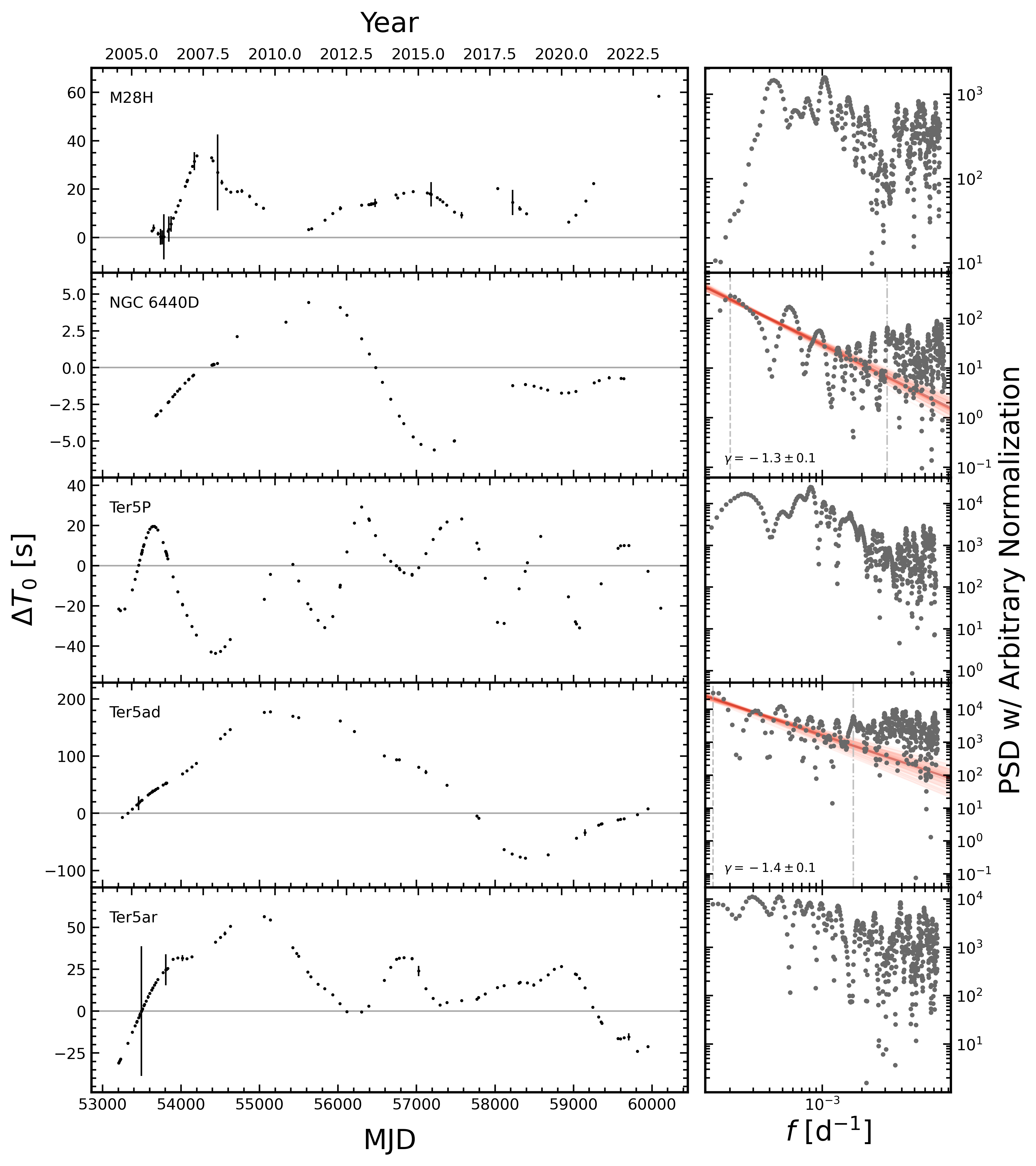}
    \caption{\textit{Left:} As in Figure \ref{fig:T0_progression}, phase variations ($\Delta T_0$) for each system over time using their final \tzx\ values.  Note that NGC 6440D and Ter5P are shown with their respective \pbdot's removed.  \textit{Right:} Over-sampled Power Spectral Densities (PSD; grey points) of the $\Delta T_0$s for each system from a Lomb-Scargle periodogram with an arbitrary normalization.  In dark red we show the best fit power-law to the frequencies between the dashed and dash-dotted lines for two systems.  The $\gamma$ values and their error for these fits are given in the bottom left of each plot.  In red, we show a multi-variate gaussian sampling representing the error region of our fit.}
    \label{fig:deltaT0}
\end{figure*}
\subsection{Isolating the MSP} \label{subsec:roi}
With more refined measurements of \tzx, we removed the short- and long-term characteristics of the binary to place each MSP in ROI.  For each group in each RB's binary-piecewise model, we first used \pint\ to fit the simple-binary model for only \tzx\ to get its uncertainty and, in a very small number of cases, updated the \tzx\ measurement to further improve its accuracy.  Next we inspected a residual vs. phase plot from a small set of TOAs fit with the simple-binary model to define an eclipse region around superior conjunction (see Table \ref{tab:eclipse_regions} for each system's region), and we removed all TOAs with orbital phases coincident with this region.  Removing this region gives us a set of non-eclipsed TOAs, \ti. 

\begin{table}
    \centering
    \caption{Region of orbital phase where TOAs are removed for each system.}
    
    \begin{tabular}{@{} lrr @{}}
    \toprule
    
    \multicolumn{1}{l}{\footnotesize{System}} & \multicolumn{1}{c}{\footnotesize{Lower Bound}} & \multicolumn{1}{r}{\footnotesize{Upper Bound}} \\
    
    \midrule
    
    M28H & 0.17 & 0.41 \\
    NGC 6440D & 0.15 & 0.35 \\
    Ter5P & 0 & 0.5 \\
    Ter5ad & 0.08 & 0.4 \\
    Ter5ar & 0.125 & 0.4 \\

    \bottomrule
    
    
    \end{tabular}
    
    \label{tab:eclipse_regions}
\end{table}

We then computed the Roemer delays ($\Delta_{R}$) at \ti\ based on the simple-binary model, and we created an identical set of TOAs, \trd, from which we then remove the Roemer delay.  We also needed to apply barycentric to topocentric corrections ($\beta_{i}$) to the predicted Roemer delays, which requires corrections to $\Delta_{R}$ for both \ti\ and \trd\ to properly describe an isolated state.  To do this we used the simple binary model to compute corrections for the $\Delta_{R}$ corresponding to \ti, $\beta_{t}$, and we used the isolated model to compute corrections for the $\Delta_{R}$ corresponding to \trd, $\beta_{\Delta_{R}}$, as these TOAs are now in a form of isolation.  Using these two terms and $\Delta_{R}$, we then computed a first-order correction that describes the barycentering effects needed to remove the orbital timing delays from each group, defined as:
\begin{equation}
    \beta_{i} = \left(\beta_{t}-\beta_{\Delta_{R}}\right) - \Delta_{R}.
\end{equation}
We finally removed $\Delta_R$ and $\beta_i$ from \ti.  The full sets of TOAs including all of the above corrections then represent the MSP in a state of ROI described by the isolated par file.  Tests using simulated binary TOAs reveal that we are able to remove orbital effects to better than 100\,ns using this technique.

\subsection{Fitting \& Inflating Uncertainties} \label{subsec:fitting}
We fit for parameters such as the pulsar spin frequency, frequency derivative, position and proper motion, using \pint.  Initially, we use \pk\ -- a GUI-based implementation of \pint -- with a Downhill weighted least-squares fitter \citep{TEMPO2,Lommen2013,Susobhanan2024} to get a preliminary, long-term fit as \pk\ allows for quick changes of various fitting parameters and TOA grouping.  We included positions, proper motion in right ascension, DM, frequency, and various numbers of spin frequency derivatives in these fits.  \citet{Susobhanan2024} performed various noise parameter tests that show \texttt{PINT}'s measured uncertainties on TOAs via fitting algorithms are likely underestimated on their own.  They also added noise parameters to the \pint\ Downhill fitter that can be used to inflate the uncertainties to better describe the measured physical quantities in the fit.  We therefore added two fitted error factors (\texttt{EFAC}s) to our models -- one for the incoherent data and the other for the coherent data -- to inflate the errors from our initial fits.  This inflation helps to compensate for the systematics due to un-measured and un-modeled DM effects caused by perturbations in the TOAs from the ionized gas in these systems.  We chose not to include a fit for errors added in quadrature (e.g., EQUAD) as it may not properly describe the white noise of the isolated TOAs since the initial datasets are dominated by systematics.  We then used \pint\ to perform the final fitting for each system, and we centered the epoch for our measurement of the spin period and the positions along our baseline.

\section{\bf{Results}} \label{sec:results}

\subsection{Pulse Profiles} \label{subsec:pulses}
With the exception of Ter5ar, which is the most recently discovered \citep[see Figure 3 in][]{Padmanabh2024}, the previously published pulse profiles for these RBs were only able to make use of the incoherent data available at the time.  Furthermore, the coherent data for the Ter5 systems in this work were not analyzed by \citet{Martsen2022} due to the general difficulty in aligning and summing RB profiles -- a process aided here by our analysis of the binary orbits.  We therefore include summed pulse profiles derived from our coherent dedispersion observations for each MSP in Figure \ref{fig:pulse_profiles}.  Each profile is a sum of many tens to over 100\,hours of S-band, coherent-dedispersion data centered at 2\,GHz.  Of note, we point out the secondary peak that can be seen in the profile of Ter5P, and we also see a noticeably sharper peak in the profile of Ter5P and Ter5ad than that of their incoherent profile in \citet{Ransom2005} and \citet{Hessels2006}.

\begin{figure*}
    \centering
    \includegraphics[width=0.9\textwidth]{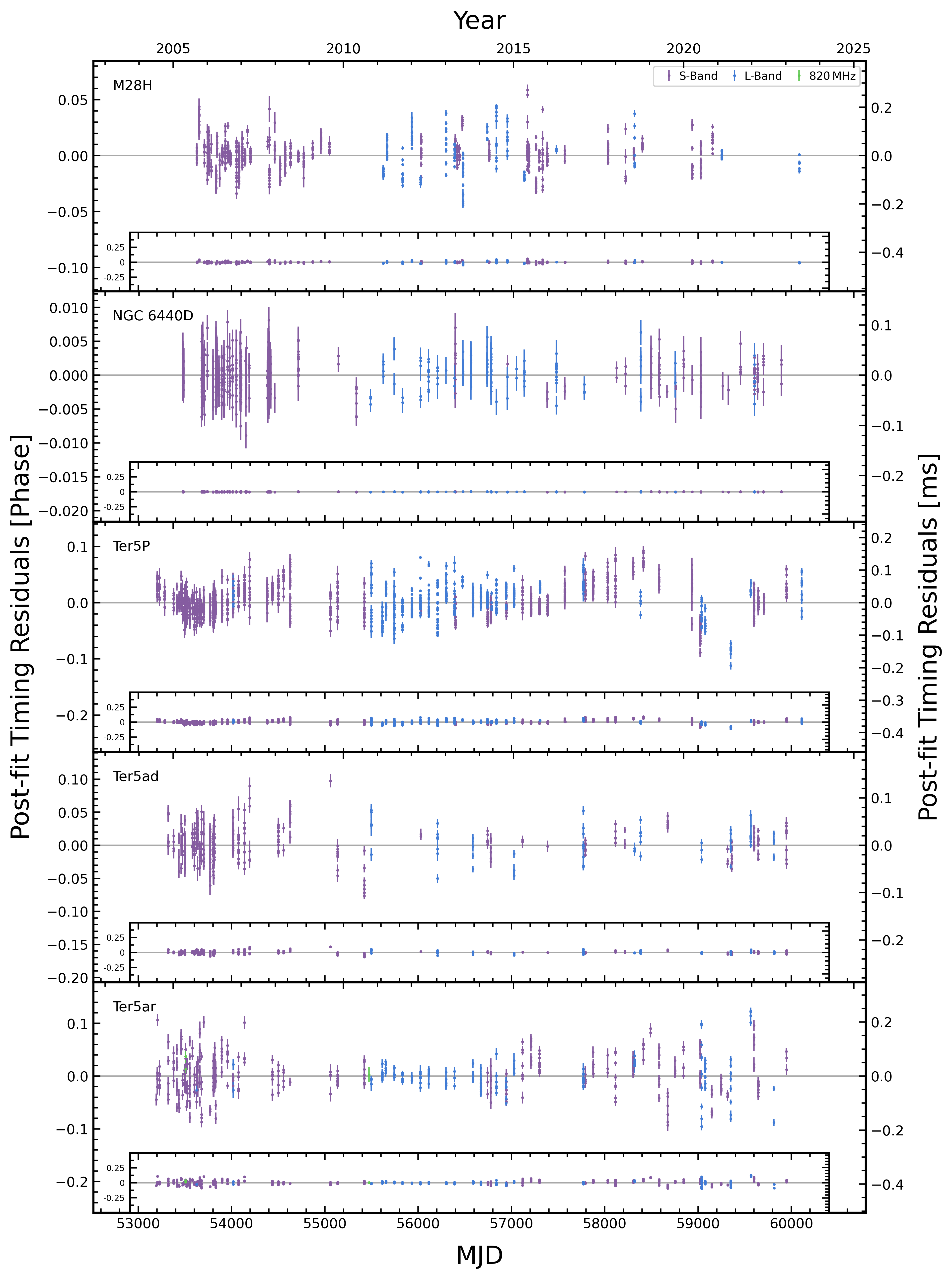}
    \caption{Timing residuals for the five RBs described in this work after applying the ROI technique.  We note that the errors shown here are our \textit{measured} uncertainties from \textsection\ref{subsec:roi}, not the inflated errors used to measure the spin properties in \textsection\ref{subsec:fitting}.  Each inset plot shows the same data, but with axes noting $\pm0.5$ pulse phase to highlight how close to zero the residuals are.}
    \label{fig:timing_solutions}
\end{figure*}

\subsection{Phase Variations} \label{subsec:phase_variations}
The orbital variations/wander in RBs are a driving force of the systematics in the long-term solutions of each system.  Thus, we show in Figure \ref{fig:deltaT0} our final measurements of \tzx\ for the five RBs and how they compare to the value predicted by each simple-binary model over the baseline of each system.  These variations represent a large portion of the orbital information removed by the ROI technique.  We also show the same variations for Ter5P and NGC 6440D with the \pbdot\ removed in Figure \ref{fig:PBdots}.  In \textsection\ref{sec:methods}, we outline how important accurate measurements of \tzx\ are to effectively understanding these variations and performing subsequent timing; therefore, we include all of the above measurements of \tzx\ in Table \ref{tab:T0s}.

We also include the power spectral density (PSD) for the $T_0$ deviations for each system computed using a Lomb-Scargle periodogram \citep{LombScargle1982,astropy:2013,astropy:2018,astropy:2022} in Figure \ref{fig:deltaT0}.  The slope of these PSDs is the spectral index, $\gamma$, of the orbital phase variations, meaning it may contain information relating to the underlying mechanisms that drive change to the orbital period.  It is clear from the PSDs that most of the systems are poorly described by a power law. 
We do, however, show fits for NGC 6440D and Ter5ad, which have behaviors more similar to a power law.  These fits take into account the frequencies between the bins partially covariant with \pbdot\ (e.g.~$\gtrsim 1/(10\,\textrm{yr})$) and the bins associated with white noise (e.g.~$\lesssim 1/(1\,\textrm{yr})$).
For reference to what the $\gamma$ values would be, the best-fit $\gamma$ and its error are shown in the bottom left of each plot. 

\subsection{Long-term Timing Solutions} \label{subsec:timing_solutions}
Timing residuals from our fully-phase-connected timing solutions for the spin behavior of each RB after performing the ROI technique and fitting are shown in Figure \ref{fig:timing_solutions}.  The resultant parameters for each system are given in Tables \ref{bintab1} \& \ref{bintab2}.  Despite the significant systematics posed by the orbit and circumbinary material, we are still able to very accurately measure the spin properties of the MSPs over nearly 20 years.

\section{\bf{Discussion}} \label{sec:discussion}
The success of the ROI technique in allowing for long-term descriptions of MSP spin behavior marks a significant step in the pursuit of timing pulsars in RBs with dramatic orbital variations.  Even in the presence of large systematics, this novel method allows us to account for each MSP's rotation over nearly 20 years.

\subsection{Applegate Model Applied to Ter5P} \label{subsec:Ter5P}
The Applegate mechanism \citep{Applegate1992} is the most common theory of how variations in a companion star can be used to describe periodic orbital variations in RBs and other compact binaries.
In the model, the gravitational quadrupole (GQ) moment of the companion star can change, perhaps due to magnetic activity, and this can couple with the orbital period and give rise to variations. While the model is often applied to other binary types, the model was created to describe Algol-type binaries.  Thus, difficulties can arise when applying it to RBs, as well as to the more well-behaved black widow systems).  Nonetheless, it has been practically used to describe the behavior of a few black widows.  For example, PSR B1957+20 has quasi-periodic oscillations that give a $\Delta P_{\rm b}/P_{\rm b}=1.6\times10^{-7}$; \citep{Arzoumanian1994,Applegate1994} arising from orbital variations on the order of a few seconds.  Similarly, PSR J2051-0827 has had several investigations into its orbital variations (on the order of $\sim$10\,s) that model the GQ of the companion \citep{Lazaridis2011,Shaifullah2016,Voisin2020}.  For RBs, GQ analysis has been carried out in radio-timing studies of the transitional MSP, PSR J1023+0038 \citep{Archibald2013}, and in $\gamma$-ray timing of PSR J2339-0533 \citep{Pletsch2015}, PSR J2039-5617 \citep{Clark2021}, PSR J0838-2827, PSR J0955-3947, and PSR J2333-5526 \citep{Thongmeearkom2024}.  Peak-to-peak, Ter5P exhibits much larger orbital variations than most of these systems, having an amplitude only smaller than PSR J0955-3947. However, the variations in PSR J0955-3947 are much more similar to those of Ter5ad in this work, appearing like steep-spectrum red noise as the \dt\ variations show up with a dominant cubic term over its 15\,yr timing solution.  This makes Ter5P somewhat unique among spiders, having both a large number of oscillations over nearly 20 years while also exhibiting consistent, large variations in radio observations.

\begin{figure*}
    \centering
        \subfloat{
         \includegraphics[width=0.49\textwidth]{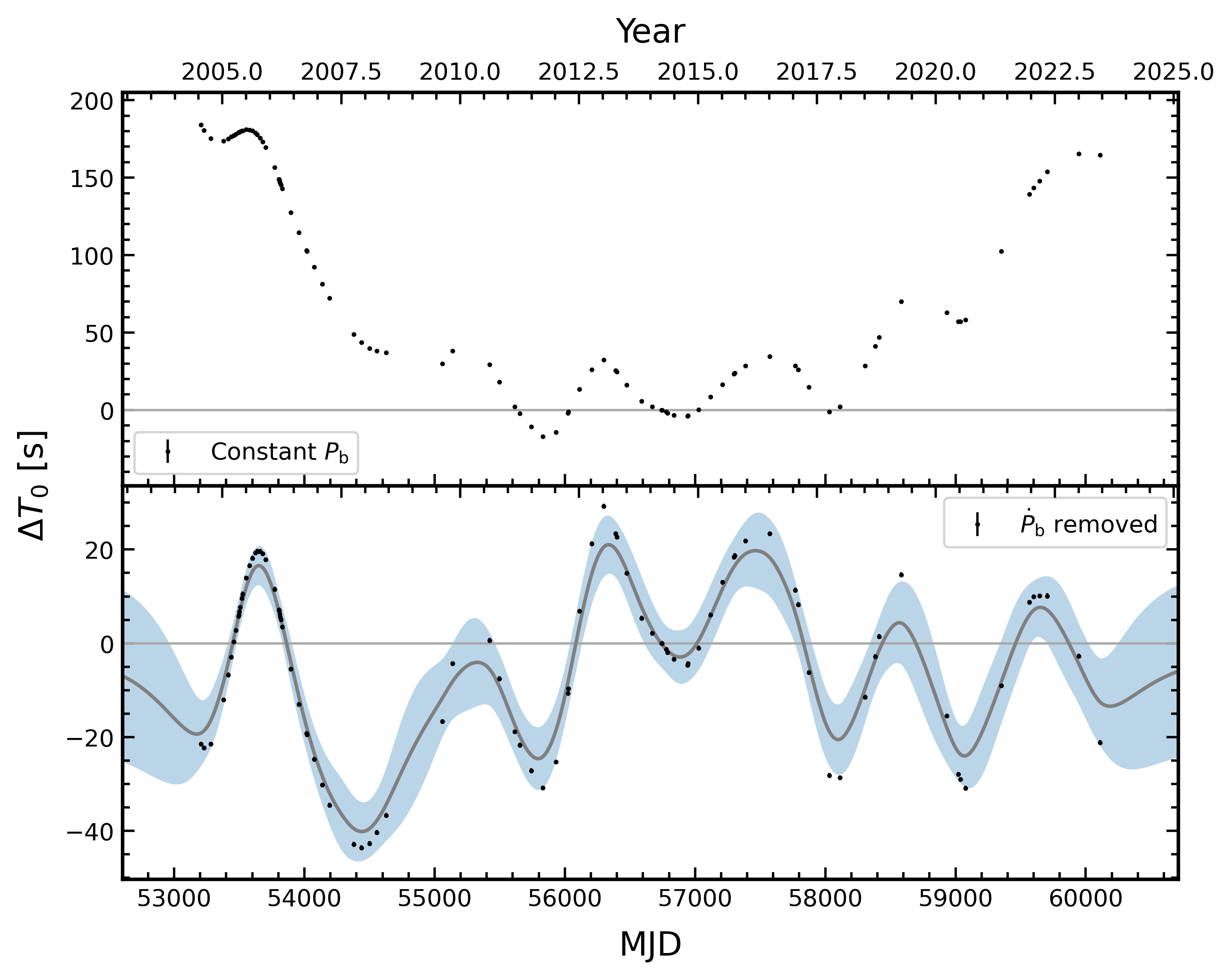}
        \label{fig:Ter5P}
        }
        \subfloat{
             \includegraphics[width=0.49\textwidth]{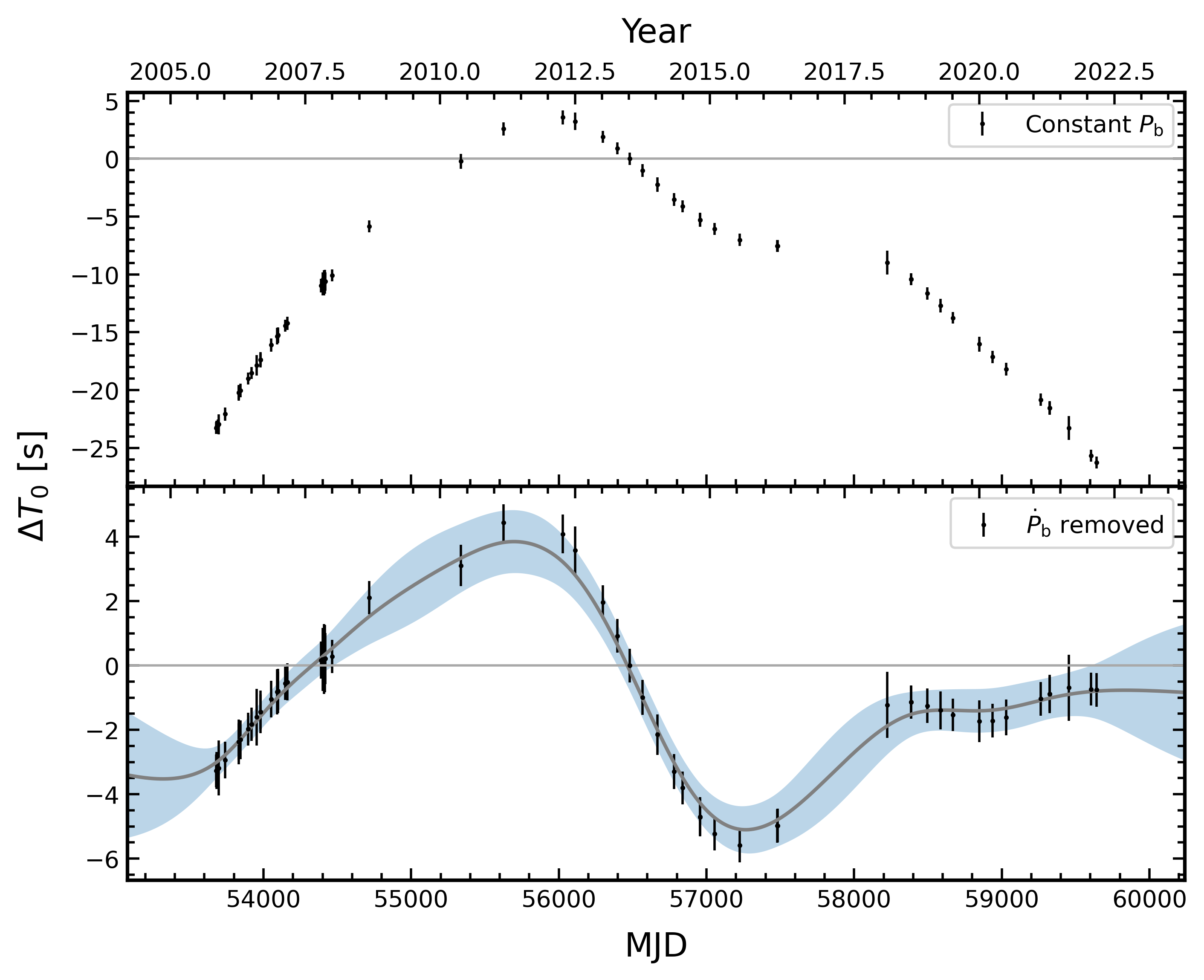}
            \label{fig:NGC6440D}
        }
    
    \caption{Phase variation ($\Delta T_0$) trends over time for Ter5P (left) and NGC 6440D (right) assuming a constant orbital period (upper panels) and after the removal of the best-fit \pbdot\ (lower panels).  Also shown in the bottom panels are Gaussian process regressions to the measured \tzx\ values (note these values match those in Figure \ref{fig:deltaT0}).  The quasi-periodic oscillations described in \textsection\ref{subsec:Ter5P} for Ter5P are evident.}
    \label{fig:PBdots}
\end{figure*}

Visual inpspection of Figure \ref{fig:Ter5P} with the \pbdot\ removed shows it could be possible that quasi-periodic oscillations consistent with the Applegate mechanism might be present in the observed variations of Ter5P.  We fit a simple sine wave to the portion of the data above MJD=57500, and from this fit we derive a semi-amplitude of 22.4\,s and $P_{\rm b,mod}$ of $\sim$1033\,d.  Using Equation 38 in \citet{Applegate1992}, this gives $\Delta P_{\rm b}/P_{\rm b}=1.6\times10^{-6}$.  However, it is clear that both the modulation amplitude and period are not constant over the baseline of our observations.  It is also plausible that there are additional or even alternative mechanisms (such as the quadrupole moment of the gravity field of the pulsar's companion) responsible for the variations that are seen here, especially since we removed a large \pbdot\ value to uncover them.  

\subsection{Timing Solutions} \label{subsec:timing_discussion}
The small dispersion in each of the timing residuals in Figure \ref{fig:timing_solutions} shows that the noise in the orbits of these system does not necessarily strongly affect the clock of the MSP's spin.  Our data are well described by solutions containing astrometry, as well as small numbers of frequency derivatives -- where derivatives after $\dot{f}$ are expected due to the accelerations of the cluster \citep{Phinney1992,Phinney1993,Freire2017,Prager2017}.  Some of the variations present in the post-fit residuals are almost certainly a manifestation of \tzx\ values that are still not yet accurate enough to describe the local solution (e.g., the TOAs around MJD$\sim$56600 in M28H).  It is important to point out that large systematics caused by other mechanisms are still present in these systems, though.  Small duration eclipses will cause issues with the TOAs by affecting the DM value in that observation \citep[e.g.][]{nt1992,Archibald2013,Nice2000}.  Infalling material can also cause a torque on the MSP's magnetosphere that will affect the spin.   These variations are clear in the timing solution for Ter5P (the most dramatic around MJD$\sim$59400), where efforts to further improve the local variations as well as additional frequency derivatives in the timing model failed to remove these sharp features.  Similar features are also seen in the long-term timing solution of Ter5A \citep{Rosenthal2024}.  A torque from the material would affect the pulsar as a second derivative change in rotational phase:
\begin{equation}
    \tau = I\frac{d^2\phi}{dt^2}.
\end{equation}
These torques are likely present at small levels in all of our timing solutions, but the exact magnitude and cause of them likely varies greatly system to system.  It is not clear whether the torques would be caused by effects from the companion or nearby stars.

\subsection{Positions and Proper Motions} \label{subsec:positions}
We fit for right ascension (\ra), declination (\dec), and the proper motion only in \ra\ ($\mu_{\alpha}$) in all analyses of the isolated TOAs (see \textsection\ref{subsec:fitting}).  Initial assessments of fits allowing proper motion in \dec\ ($\mu_{\delta}$) to vary yield non-physical measurements (e.g., too large for the pulsar to remain in the GC), likely due to the low ecliptic latitudes $|\beta| \lesssim 3\degree$ of these globular clusters, where pulsar timing has much less power measuring declination; therefore, we include $\mu_{\delta}$ as a fixed value of 0 in our fits.  The sensitivity to \dec\ and by extension $\mu_{\delta}$ in fits of long baseline TOAs are inherently less precise than those of \ra\ for these pulsars, and the large systematics imposed by the phase variations compound to make measurements of $\mu_{\delta}$ exceedingly difficult.  Future work and additional observations could seek to improve upon this.  It is also important to note that the positions of the pulsars analyzed here were already precisely measured, so only one iteration of the ROI technique was necessary.  When using this technique to time newer and slower pulsars, the process of the ROI technique would need to be iterated if there is a significant position shift in the timing results to redo the Solar System delay subtraction.

\begin{figure*}
    \centering
    \includegraphics[width=\linewidth]{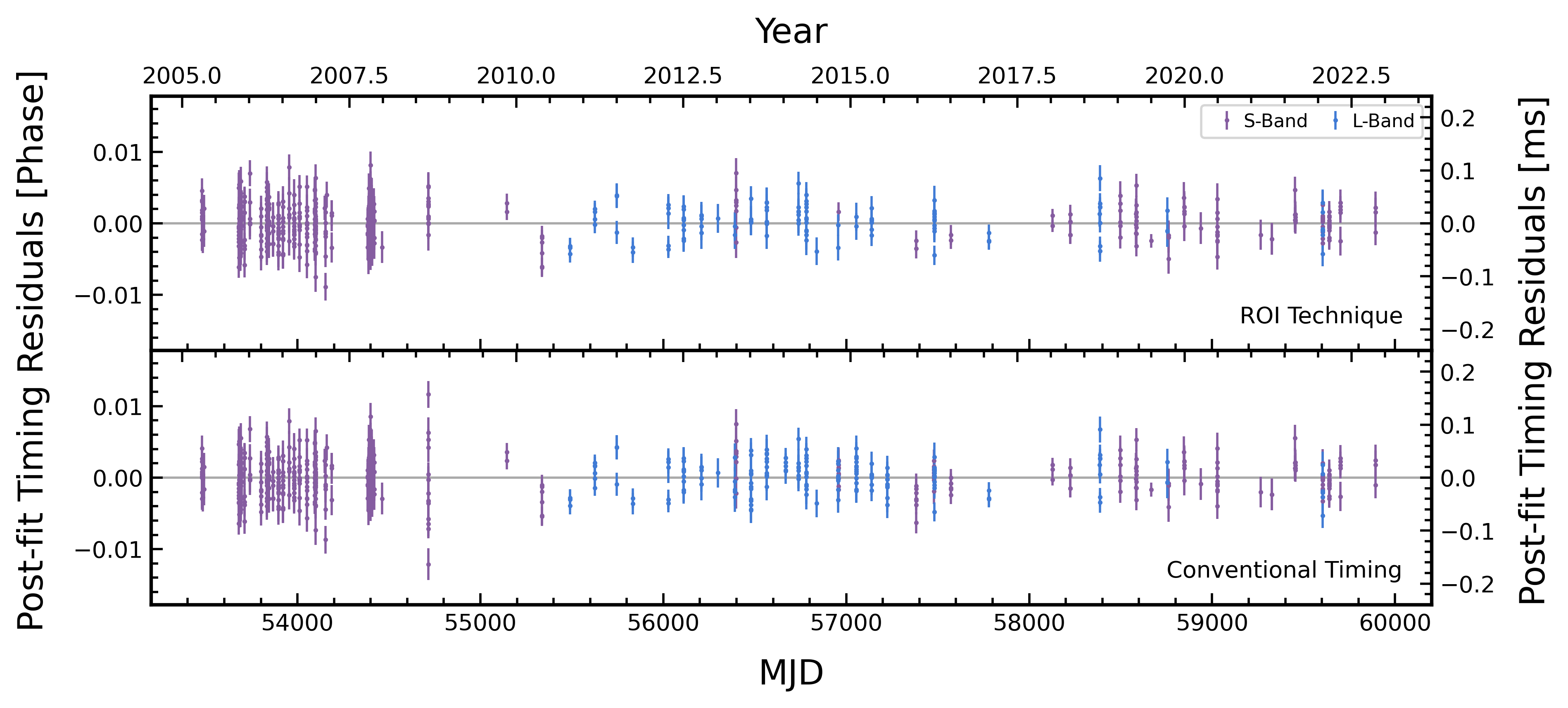}
    \caption{Comparison of timing residuals obtained via the ROI technique (top) and those obtained using conventional timing techniques (bottom).  It is clear that these two methods yield nearly identical results.}
    \label{fig:timing_comparison}
\end{figure*}

As a consistency check, we compare our positions to those of both radio and X-ray positions for each system where available.  M28H has both radio and X-ray \citep{Vurgun2022} positions.  We cannot draw any conclusions on the consistency with the cited radio positions as the reported uncertainties are too small to be feasibly correct\footnote{This is likely due to typographical errors in the formatting of the radio coordinates in Table A1.}; however, our \ra\ and \dec\ differ from the reported values of  $\alpha=18$:24:31.61052125 \& $\delta=-$24:52:17.2268378 by only 0.7 and 760\,mas, respectively.  Our values are consistent within errors with the reported X-ray positions. 
\citet{Freire2008} reports radio-timing positions of $\alpha=17$:48:51.64665(7) and $\delta=-$20:21:07.414(18) for NGC 6440D, and our \dec\ value is entirely consistent with that result.  Our \ra\ is not consistent within errors, but as was the case with M28H, our value differs by only 0.95\,mas.  Some of the systems in Ter5 have radio positions from \citet{Urquhart2020}.  Our measurement of both \ra\ and \dec\ for Ter5P and Ter5ar are consistent within errors with the reported values for sources VLA5 and VLA38, in that work, respectively.
Ter5P and Ter5ad have X-ray postions from \citet{Bogdanov2021} that are consistent within errors with our measured values. \citet{Bahramian2020} report positions for CXOU J174804.63$-$244645.2, which \citet{Padmanabh2024} identified as being consistent with Ter5ar.  The positions we measure for Ter5ar are not consistent within error with the reported values of $\alpha=17$:48:04.63(13) and $\delta=-$24:46:45.34(13); however, the uncertainties are based off of source extraction via centroiding that may not represent the uncertainties of a dedicated measurement of the X-ray position.  Given that our measurements differ by only 10 and 530\,mas in \ra\ and \dec, respectively, it is likely these values would be consistent with uncertainties produced with \citeauthor{Hong2005}'s \citeyearpar{Hong2005} expression for 95\% confidence error circles such as those in \citet{Bogdanov2021}.

Similarly, we compared our measured $\mu_{\alpha}$ to those measured for each GC by \citet{Vasiliev2021} using \textit{Gaia}.  The reported values for M28, NGC 6440D, and Ter5 are $-0.278\pm0.028$, $-1.187\pm0.036$, $-1.989\pm0.068$\,mas\,yr$^{-1}$, respectively.  As shown in Tables \ref{bintab1} \& \ref{bintab2}, our measurements for NGC 6440D and Ter5ad are consistent within errors with the reported values, while our measurements for M28H, Ter5P, and Ter5ar are not.  These values are relatively reasonable given the proper motions of stars in the cluster within an order of magnitude or two, though, and our measurement for Ter5P only differs by 0.189\,mas\,yr$^{-1}$.  As with $\mu_{\delta}$, our measurements of $\mu_{\alpha}$ are likely substantially affected by the systematics of the phase variations.  The relative closeness of our measurements are a sign that improvements to the ROI technique in the future may yield accurate $\mu_{\alpha}$ measurements.

\vfill
\subsection{ROI Timing vs. Traditional Timing} \label{subsec:timing_comparisons}
In Figure \ref{fig:timing_comparison} we show a comparison of our long-term timing solution for NGC 6440D created via ROI and the long-term timing solution from Ransom et al. (in prep).  The latter solution contains nine orbital frequency derivatives conventionally used for timing.  NGC 6440D is by many measures the most well-behaved of the systems in our sample, having orbital variations of only a few seconds, typically.  The results of both methods look to be nearly identical, which is a useful check on the efficacy of the ROI technique.  

We stress that this is not to say that conventional timing techniques should not be used to derive long-term solutions for RBs.  Rather we highlight that for systems with significant orbital variations there are useful benefits that our technique provides over conventional methods.  Conventional-timing techniques necessitate \pb\ and nine orbital frequency derivatives to describe the binary effects of NGC 6440D and the effects of the cluster, whereas the ROI technique produces nearly identical results with a simple model that remove binary information entirely.  Similarly, higher-order, orbital frequency derivatives poorly predict orbital behavior in large gaps between observations, while the stable spin of the MSP -- the connecting factor in our method -- extrapolates nicely over those same gaps.  Thus, the ROI technique produces nearly identical results by trading orbital frequency derivatives for a number of individual, well-measured \tzx\ values needed to accurately describe the long-term timing behavior of the systems.

\begin{figure*}
    \centering
    \includegraphics[width=\linewidth]{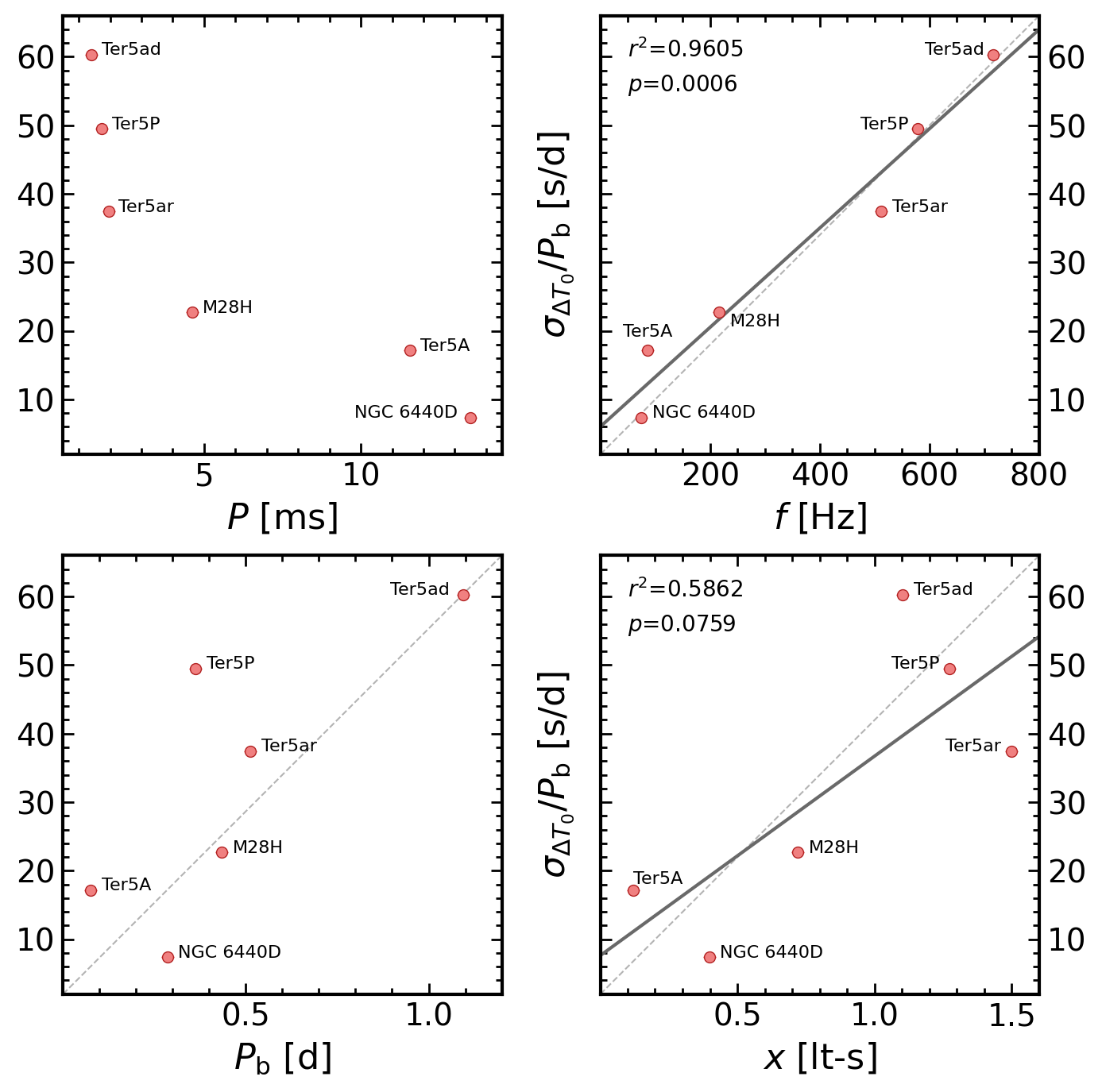}
    \caption{
    The fractional deviation of the five RBs described here, as well as Ter5A, plotted against spin period (upper left), spin frequency (upper right), orbital period (bottom left), and projected semi-major axis (bottom right).  Values for Ter5A come from analysis presented in \citet{Rosenthal2024}.  Also plotted as a solid line is a linear regression fit for the strong correlation seen with spin frequency and the weaker one with projected semi-major axis, as well as the regression's $r^2$ and $p$-values.  The light-gray, dashed lines in each plot show the one-to-one line for that variable.}
    \label{fig:covariance_inspection}
\end{figure*}
\subsection{Phase Variation Analyses} \label{subsec:covariances}

We inspected the standard deviation of the phase variations ($\sigma_{\Delta T_0}$) as a fraction of the orbital period.  This experiment was to see if, fractionally, the observed variations were similar or if some systems have higher variations than others.  In the four plots of Figure \ref{fig:covariance_inspection}, we show the fractional deviation ($\sigma_{\Delta T_0}/P_{b}$) plotted against spin frequency and period, \pb, and semi-major axis.  Immediately evident was the correlation $\sigma_{\Delta T_0}/P_{b}$ has with spin frequency (and inversely with spin period) and potentially with the projected semi-major axis.  Although there is more noise in the trend, it is clear that, as the binary distance tightens, the fractional deviation in the phase variations becomes less drastic.  The near direct relationship with spin frequency is certainly the most striking.

It is unclear what the physical mechanism for this relationship would be.  Younger MSPs that have just been spun up by their binary companion may have the fastest spins and be closer to the point where they overflowed their Roche lobe to contribute the material for spinning up the pulsar.  This could mean more material would be present to perturb the system, more interactions would be taking place as the pulsar finishes its spin up, or both.  Older and slower MSPs would be then be farther from this evolutionary stage and more settled, and thus may not be prone to large variations.  
Owing to the precise nature of the ROI technique, it is possible that this is a manifestation of the sensitivity of our \tzx\ measurements, where the more rapidly a system is rotating, the more accurate the values of \tzx\ need to be.  Future long-term studies of these systems may be able to disentangle the physical reasoning behind these strong trends, assuming they are not byproducts of small statistics.  

\subsection{Mitigating Systematics in the Future} \label{subsec:systematics}
As has been mentioned throughout the text, there are systematics present in these datasets that serve as contaminants.  We are able to remove a good portion of them through the ROI technique; however, it is not to be assumed that all systematics have been removed at this point.  While it is likely not possible to account for every variation caused by the eclipses, gas in the system, or any torques imposed by accreted material, finding more ways to even further improve the \tzx\ measurement for a piecewise group can in fact go a long way to mitigate local systematics.  One simple addition to the ROI technique for future studies could be to implement a simple routine that finds a \tzx\ that both minimizes the dispersion of TOAs in post-fit phase and keeps the group close to a long-term trend line of the isolated TOAs.  We also reiterate that our analysis was conducted using time-series data, as it keeps the problem at hand far simpler to address.  Future studies will be able to leverage the multi-frequency information of raw data to account for DM variations and delays in certain observations.  These variations are likely some of the largest contaminants to measurements of \tzx, and our analysis of the time-series data only removes a small fraction of them by removing TOAs near the eclipse.  

\section{\bf{Summary}} \label{sec:summary}
We have presented a novel technique for ``isolating'' the underlying MSP clock in binaries with dramatic orbital variations by removing the timing delays from the binary orbit and fitting local variations in the TOAs.  This technique allows for phase connection over baselines that far surpass those of the initial solutions containing binary information.  We used this technique to get timing solutions for five RB systems found in three GCs spanning almost two decades.  The results of our solutions derived from isolated TOAs are consistent with those derived from conventional timing techniques for GC pulsars, meaning this is an effective alternative for systems that show large phase variations.  Variations seen in the five systems we investigated show the broad spectrum of systematics present in RBs, ranging from relatively well behaved (e.g., NGC 6440D) to quasi-periodic (e.g, Ter5P) to unpredictable (e.g, M28H \& Ter5ar).  An analysis of Ter5P shows that it is possible that its oscillations arise from the Appelgate mechanism; however, the changing nature of the oscillations may necessitate additional or alternative mechanisms to fully describe the phase variations.  A striking correlation exists between the standard deviation of a system's variations as a fraction of its orbital period and its spin frequency.  Whether this is a probe of the MSPs age since being spun up, the inherent susceptibility of an MSP to perturbations based on its spin, or even just a manifestation of the need for duly accurate measurements of phase, it is clear that future studies of RBs will need to continue this investigation.  The nature of RB timing is riddled with pervasive systematics that limit our ability to describe these systems in ways other MSP binaries can be over similar baselines.  With refinement to the ROI technique, multi-frequency information, and more systems with long baselines, though, it is still possible to investigate interesting physics problems with these unique binaries.

\section*{\textbf{Acknowledgements}}
The Green Bank Observatory is a facility of the National Science Foundation operated under cooperative agreement by Associated Universities, Inc.
The National Radio Astronomy Observatory is a facility of the National Science Foundation operated under cooperative agreement by Associated Universities, Inc.

Support for this work was provided to K.A.C. by the NSF
through the Grote Reber Fellowship Program administered by Associated Universities, Inc./National Radio Astronomy Observatory. S.M.R. is a CIFAR Fellow and is supported by the NSF Physics Frontiers Center awards 2020265.  Pulsar research at UBC is supported by an NSERC Discovery Grant and by the Canadian Institute for Advanced Research.

We would like to acknowledge Brian Prager for his part in presenting some of the equations found in Appendix \ref{sec:appendixdynamics}.

%





\bibliography{references}{}
\bibliographystyle{aasjournal}





\newpage
\appendix
\section{Dynamical Considerations}\label{sec:appendixdynamics}
In this work, we have assumed that the dominant systematic in our timing analysis is changes to the time of periastron passage (i.e., $\Delta T_0$).  Empirically, this assumption yields good results, and the addition of DM effect modeling in future implementations will further improve the method's performance.  However, from a dynamics perspective, it is unclear if $\Delta T_0$ can be the sole orbital element responsible for observed timing variations.  In the text that follows, we outline some dynamical considerations that could be useful in investigating this further.  For additional commentary detailing how some equations here relate to studies of MSPs in GCs, see \citet{Prager2017PhDT} when they are originally presented.

\subsection{Relating $T_0$ to $P_{\rm b}$ \& $\dot{P}_{\rm b}$}
For investigations of the orbital period, we relate changes in orbital phase to changes in $T_0$ as follows: 
\begin{equation}\label{eqn:dphi}
    \Delta\phi = n\Delta T_0 = \frac{2\pi}{P_{\rm b}}\Delta T_0,
\end{equation}
where $n$ is the orbital frequency ($n=2\pi/P_{\rm b}$).  When considering a changing orbital period, we can similarly express the change in orbital phase in terms of:
\begin{equation}
    \Delta\phi = \frac{1}{2}\dot{n}\Delta t^{2} = \frac{-\pi\dot{P}_{\rm b}}{P^{2}_{\rm b}}\left(\Delta t\right)^{2},
\end{equation}
which relates relates the characteristic change in $P_{\rm b}$ to the observed $T_0$ wander over the timescale $\Delta t$ through
\begin{equation}
    \frac{\Delta P_{\rm b}}{P_{\rm b}} \sim 2\frac{\Delta T_{0}}{\Delta t}.
\end{equation}
The characteristic orbital period derivative is then:
\begin{equation}\label{eqn:pbdott0}
    \dot{P}_{\rm b} \simeq \frac{\Delta P_{\rm b}}{\Delta t} = 2 P_{\rm b} \frac{\Delta T_{0}}{\left(\Delta t\right)^{2}}.
\end{equation}

\subsection{Relating $T_0$ to $x$}
For our analyses, we assumed that the semi-major axis is constant over the duration of our baseline, attributing all orbital changes to the time of periastron passage, $T_0$.  For investigations interested in exploring this assumption, we give a relationship between the $T_0$ and the projected-semi-major axis, $x$.  From Kepler's third law, the shrinking of the pulsar orbit by gravitational wave damping relates the change in $x$ to $P_{\rm b}$ as:
\begin{equation}
\frac{\dot{x}}{x} = \frac{2}{3}\frac{\dot{P_{\rm b}}}{P_{\rm b}}.
\end{equation}
Replacing $\dot{P}_{\rm b}$ with the result of Equation \ref{eqn:pbdott0}, this becomes:
\begin{equation}
\dot{x} = \frac{4}{3} x \frac{\Delta T_0}{\left(\Delta t\right)^2}.
\end{equation}
Similarly to $\dot{P}_{\rm b}$, the characteristic change in $x$ is then:
\begin{equation} \label{eqn:dotx}
\dot{x} = \frac{4}{3} x \frac{\Delta T_0}{\left(\Delta t\right)^2} \simeq \frac{\Delta x}{\Delta t}.
\end{equation}
Rearranging Equation \ref{eqn:dotx}, we get the relationship:
\begin{equation}
\Delta T_0 \simeq \frac{3}{4} \frac{\Delta t}{x} \Delta x.
\end{equation}

\section{Example ``isolating'' code}\label{appendix:code}

The following is an example Python script that shows that {\tt PINT} commands that could be used to ``isolate'' TOAs from a binary pulsar where T0 has been measured for each observation.

\begin{lstlisting}[language=Python]
import copy
from pint.models import get_model_and_toas, get_model
import pint.logging
pint.logging.setup(level="WARNING")

def get_T0_for_MJD(MJD):
    """This code returns a best T0 for the pulsar on a given MJD"""
    # This should be written by the user, based on how
    # correct T0s have been measured and stored.
    return T0

# Read in the simple binary model for the pulsar and the original TOAS
mbin, tbin = get_model_and_toas("PSR_binary.par", "PSR_binary.tim")
# Set our long-term "reference" values for PB and T0
PB, T0ref = mbin.PB.value, mbin.T0.value
# This will allow us to explicitly compute the Roemer delay later
bc = mbin.get_components_by_category()["pulsar_system"][0]
# Load a different parfile where all binary lines have been removed
miso = get_model("PSR_isolated.par")
# Separate the TOAs into observations on different days
obss = tbin.get_clusters()
# This is the number of different observations
Nobss = obss.max()

# Loop over the observations to produce a set of *isolated*.tim files
# which can be concatenated and used for timing the "isolated" pulsar
for ii in range(Nobss):
    # Make a mask for the TOAs for the observation in question
    tmask = obss==ii
    # Select our TOAs for this day
    tmasked = tbin[tmask]
    # Get our best T0 value for this observation
    T0 = get_T0_for_MJD(tmasked.get_mjds().value.mean())
    # Update the orbital values to make parfiles for each observation
    mbin.T0.value = T0
    mbin.PB.value = PB + mbin.PBDOT.value * (T0 - T0ref)
    # Write out the parfile for the observation
    mbin.write_parfile(f"PSR_binary_{T0:.2f}.par")
    # Write out the binary TOAs for the observation
    tmasked.write_TOA_file(f"PSR_binary_{T0:.2f}.tim", include_pn=False)
    # Compute the Roemer delay for the TOAs
    roemer = bc.binarymodel_delay(tmasked)
    # Copy and subtract off the Roemer delay from the original TOAs
    ttmp = copy.deepcopy(tmasked)
    ttmp.adjust_TOAs(-roemer)
    # Compute barycentric TOAs from both sets of TOAs with both models
    btold = mbin.get_barycentric_toas(tmasked)
    btnew = miso.get_barycentric_toas(ttmp)
    # Compute a first-order correction for the barycentering effects
    corr = (btold - btnew) - roemer
    # Now copy, correct, and write the isolated TOAs for the observation
    tiso = copy.deepcopy(tmasked)
    tiso.adjust_TOAs(-roemer + corr)
    tiso.write_TOA_file(f"PSR_isolated_{T0:.2f}.tim", include_pn=False)
\end{lstlisting}

\section{Supplementary Tables}\label{sec:appendixTables}
Included here are tables for data and fits used throughout the text. In Table \ref{tab:T0s} we provide the \tzx\ values used for our binary-piecewise groups.  These values are by no means final for any one group, as improvement in the accuracy of these values is an ever-ongoing process.  They should be taken as a more-refined starting place for future analyses of these sytems.  In Tables \ref{bintab1} and \ref{bintab2} we provide all of the binary information used to remove the orbital timing delays from each system as well as the parameters from out timing models of the spin behaviors.

 \newpage

\begin{longtable}{clllll}
    
    \caption{The final measurements of $T_{0,x}$ used in constructing the piecewise-continuous groups for each system.\label{tab:T0s}}\\
    
    \toprule

    \multicolumn{1}{c}{{Group}} &
    \multicolumn{1}{c}{{M28H}} &
    \multicolumn{1}{c}{{NGC 6440D}} &
    \multicolumn{1}{c}{{Ter5P}} &
    \multicolumn{1}{c}{{Ter5ad}} &
    \multicolumn{1}{c}{{Ter5ar}} \\

    \multicolumn{1}{c}{\scriptsize{\#}} &
    \multicolumn{1}{c}{\scriptsize{$T_0$ [MJD]}} &
    \multicolumn{1}{c}{\scriptsize{$T_0$ [MJD]}} &
    \multicolumn{1}{c}{\scriptsize{$T_0$ [MJD]}} &
    \multicolumn{1}{c}{\scriptsize{$T_0$ [MJD]}} &
    \multicolumn{1}{c}{\scriptsize{$T_0$ [MJD]}} \\

    \midrule

        1 & 53629.9385207(54) & 53478.316129(25) & 53204.18393645(17) & 53252.23546037(89) & 53205.2380510(25) \\ 
	2 & 53651.254881(15) & 53483.179297(65) & 53227.39148316(39) & 53320.0901308(32) & 53206.26472831(87) \\ 
	3 & 53707.8084219(91) & 53489.47281(14) & 53282.1468202(12) & 53379.189373(13) & 53216.01816206(60) \\ 
	4 & 53739.130384(38) & 53679.99454845(94) & 53378.9659555(14) & 53435.005329(14) & 53227.3116125(11) \\ 
	5 & 53753.9213147(12) & 53681.7109605(15) & 53415.22782516(67) & 53457.98837(14) & 53229.3649678(18) \\ 
	6 & 53755.66142441(40) & 53681.9970292(34) & 53439.1606633(14) & 53474.404824(18) & 53320.22591517(53) \\ 
	7 & 53756.9665067(91) & 53683.7134413(18) & 53459.1046954(14) & 53495.1990014(10) & 53379.7732062(12) \\ 
	8 & 53757.8365616(17) & 53685.7159220(16) & 53475.0599204(36) & 53500.6711532(16) & 53414.68023902(50) \\ 
	9 & 53759.141644(34) & 53687.718403(22) & 53493.55347757(39) & 53579.4701322(44) & 53434.7004489(65) \\ 
	10 & 53781.76307(11) & 53690.8651583(19) & 53496.0918081(11) & 53601.3587350(14) & 53438.80715862(55) \\ 
	11 & 53832.6613151(60) & 53696.0143945(81) & 53500.44323289(24) & 53625.4361957(17) & 53458.82736894(93) \\ 
	12 & 53844.407066(61) & 53709.745691(40) & 53506.60775125(21) & 53637.474927(36) & 53474.227529(17) \\ 
	13 & 53872.683875(37) & 53709.745691(37) & 53520.02464341(23) & 53637.474927(18) & 53493.73439988(96) \\ 
	14 & 53902.26577032(74) & 53740.6411085(32) & 53526.55177969(20) & 53659.3635263(19) & 53496.30109(45) \\ 
	15 & 53932.28269427(72) & 53801.287669(62) & 53553.38555898(27) & 53679.0632656(20) & 53501.43448085(78) \\ 
	16 & 53962.2996186(11) & 53801.573737(28) & 53579.8567112(11) & 53703.1407239(22) & 53506.5678671(16) \\ 
	17 & 53987.96626479(88) & 53833.3273607(56) & 53601.25119817(40) & 53769.9009444(12) & 53507.5945451(28) \\ 
	18 & 54052.78542310(77) & 53833.6134294(31) & 53625.18400663(33) & 53803.8282678(13) & 53526.5880762(13) \\ 
	19 & 54074.9718466(11) & 53843.3397645(39) & 53638.6008805(18) & 53804.9226976(22) & 53533.77481757(78) \\ 
	20 & 54079.75715425(75) & 53865.367053(39) & 53659.63273154(84) & 53806.017127(11) & 53554.30836398(40) \\ 
	21 & 54114.5593874(23) & 53895.11820(34) & 53679.57672240(44) & 53810.394847(29) & 53579.461957(13) \\ 
	22 & 54144.57631255(78) & 53895.4042642(11) & 53703.50950421(20) & 53814.7725655(12) & 53602.0488562(37) \\ 
	23 & 54169.807929(43) & 53920.2922392(15) & 53769.86854919(16) & 54016.1476506(15) & 53625.14909317(57) \\ 
	24 & 54202.4350143(10) & 53955.1926179(84) & 53803.59198568(20) & 54074.1524401(18) & 53625.66243175(64) \\ 
	25 & 54386.88664719(84) & 53980.9387988(51) & 53804.67983887(39) & 54138.7238191(21) & 53639.0092353(12) \\ 
	26 & 54405.1577859(65) & 53981.2248674(25) & 53805.76769160(20) & 54194.5397633(38) & 53660.05611565(62) \\ 
	27 & 54465.19152(18) & 54010.976010(14) & 53811.20695520(65) & 54500.9803277(11) & 53679.04964135(67) \\ 
	28 & 54520.004932(11) & 54010.97601(16) & 53815.19574799(34) & 54556.796287(22) & 53703.17655098(73) \\ 
	29 & 54574.38333165(93) & 54050.7395554(32) & 53819.54715879(16) & 54625.745400(11) & 53769.9105473(44) \\ 
	30 & 54632.2419688(18) & 54093.6498562(58) & 53830.78830373(76) & 55058.0451269(12) & 53804.8175548(19) \\ 
	31 & 54715.3322170(31) & 54095.6523369(35) & 53896.42208417(51) & 55136.8440094(81) & 53805.3308931(80) \\ 
	32 & 54773.1908713(95) & 54098.7990923(55) & 53957.34184827(28) & 55423.5842712(30) & 53805.84423(11) \\ 
	33 & 54871.0720237(86) & 54099.6572983(38) & 54016.08591876(19) & 55496.9109732(70) & 53811.4909528(27) \\ 
	34 & 54957.6424492(38) & 54099.657298(14) & 54018.62424305(94) & 56028.803303(13) & 53815.59765935(46) \\ 
	35 & 55048.5631694(75) & 54148.5750408(15) & 54074.46738141(38) & 56207.1949872(16) & 53819.70436542(79) \\ 
	36 & 55625.4094764(43) & 54153.438208(24) & 54138.65073591(29) & 56588.0557186(48) & 53829.97113088(95) \\ 
	37 & 55664.56195259(61) & 54160.5899248(34) & 54194.49388733(47) & 56745.6533882(18) & 53897.2184807(26) \\ 
	38 & 55833.7876745(14) & 54187.480380(34) & 54380.15431237(37) & 56778.4862517(19) & 53957.27904429(98) \\ 
	39 & 55932.5389399(19) & 54383.723481(18) & 54438.53583792(23) & 57026.9214416(40) & 54016.31291936(65) \\ 
	40 & 56026.5048941(99) & 54388.872716(97) & 54500.90618449(26) & 57118.853363(42) & 54017.339595(20) \\ 
	41 & 56032.5952797(13) & 54390.8751970(34) & 54557.11203850(57) & 57388.0825864(73) & 54074.8334548(22) \\ 
	42 & 56298.8320989(10) & 54398.8851194(28) & 54626.37216650(15) & 57770.0376157(17) & 54139.00072730(58) \\ 
	43 & 56390.62289484(77) & 54402.0318746(98) & 55057.88815504(46) & 57791.9261454(15) & 54438.7902578(11) \\ 
	44 & 56397.5833346(12) & 54404.606492(35) & 55136.93909113(38) & 58113.6875807(49) & 54500.9041975(31) \\ 
	45 & 56418.4646542(21) & 54406.0368357(75) & 55423.04502125(36) & 58215.469372(10) & 54557.371412(14) \\ 
	46 & 56421.5098467(85) & 54408.897522(11) & 55496.65645520(22) & 58320.5344765(64) & 54627.18543776(44) \\ 
	47 & 56443.2612220(43) & 54416.907445(11) & 55614.50729843(35) & 58385.1057517(27) & 55058.389480(10) \\ 
	48 & 56471.102983(20) & 54419.7681312(71) & 55653.30743345(13) & 58676.2238832(41) & 55137.95685828(45) \\ 
	49 & 56479.36850580(72) & 54463.5366351(17) & 55743.23673378(17) & 59037.3857265(58) & 55423.3726304(14) \\ 
	50 & 56738.6449081(70) & 54714.9909962(15) & 55829.90249209(13) & 59146.828722(66) & 55473.16638388(77) \\ 
	51 & 56759.5262106(42) & 55144.666093(59) & 55931.435715350(52) & 59317.5597674(69) & 55496.7799150(69) \\ 
	52 & 56837.39614888(80) & 55336.3320909(47) & 56023.178350109(53) & 59351.4870831(14) & 55614.8475620(14) \\ 
	53 & 56956.1586525(39) & 55489.950956(25) & 56028.25501932(28) & 59362.4313778(11) & 55653.3478840(38) \\ 
	54 & 57137.5650962(27) & 55625.5474924(31) & 56113.107928159(62) & 59565.9952139(16) & 55744.2086697(10) \\ 
	55 & 57172.36728814(36) & 55743.979907(30) & 56206.663658990(57) & 59599.9225153(11) & 55829.93609672(77) \\ 
	56 & 57186.2881646(17) & 55833.805455(24) & 56299.49408154(13) & 59644.7941078(12) & 55931.5769912(83) \\ 
	57 & 57187.158219(59) & 56026.3296321(39) & 56389.42339918(15) & 59814.4306563(16) & 56023.46444506(70) \\ 
	58 & 57261.98292717(47) & 56111.2920066(65) & 56399.21408991(27) & 59947.9510930(19) & 56112.78521260(35) \\ 
	59 & 57297.2201417(17) & 56207.697123(23) & 56474.2760313(14) &   & 56299.6402657(27) \\ 
	60 & 57333.76243713(41) & 56298.3808647(18) & 56587.77551551(17) &   & 56389.4744682(18) \\ 
	61 & 57382.9205236(13) & 56390.494948(27) & 56671.54035858(18) &   & 56587.62313967(64) \\ 
	62 & 57482.97680593(53) & 56397.3605938(19) & 56742.25094883(10) &   & 56671.29733409(34) \\ 
	63 & 57574.332557(16) & 56479.1762070(18) & 56745.151897270(95) &   & 56745.2180710(28) \\ 
	64 & 58034.5917351(20) & 56565.8549853(24) & 56778.150170296(93) &   & 56778.07171593(69) \\ 
	65 & 58224.263641(60) & 56668.5536037(44) & 56787.2156265(31) &   & 56838.1322731(19) \\ 
	66 & 58313.009213(11) & 56738.354336(18) & 56837.98220724(13) &   & 56941.8265555(32) \\ 
	67 & 58320.8397053(16) & 56781.2646226(25) & 56941.691102344(90) &   & 56943.8799059(11) \\ 
	68 & 58401.7547898(19) & 56837.0479954(15) & 56943.86681720(26) &   & 57026.527250(36) \\ 
	69 & 58937.2735506(36) & 56955.7664570(40) & 57026.54389180(17) &   & 57119.44131923(56) \\ 
	70 & 59029.4994048(23) & 56957.7689371(31) & 57119.37433252(49) &   & 57209.2754125(11) \\ 
	71 & 59156.96251753(89) & 57055.6043941(16) & 57209.30382670(15) &   & 57299.62286626(52) \\ 
	72 & 59255.71383447(71) & 57137.133943(18) & 57299.595923627(66) &   & 57387.4036938(42) \\ 
	73 & 60084.4415596(26) & 57224.0987970(19) & 57306.12306245(14) &   & 57573.2320873(19) \\ 
	74 &   & 57382.580804(12) & 57387.712290037(85) &   & 57769.32723824(31) \\ 
	75 &   & 57480.4162691(17) & 57573.01043610(22) &   & 57791.400786(10) \\ 
	76 &   & 57482.4187494(17) & 57769.54961759(21) &   & 57875.0749152(29) \\ 
	77 &   & 57571.958225(19) & 57791.66932139(23) &   & 58030.6163945(21) \\ 
	78 &   & 57779.357971(28) & 57875.07145397(13) &   & 58113.77717324(39) \\ 
	79 &   & 58127.503465(12) & 58030.99724588(19) &   & 58305.2522899(10) \\ 
	80 &   & 58224.480718(10) & 58113.67431029(38) &   & 58320.1390987(16) \\ 
	81 &   & 58385.8233995(14) & 58305.1372071(12) &   & 58412.5399472(14) \\ 
	82 &   & 58386.9676739(16) & 58384.9134355(16) &   & 58491.593995(11) \\ 
	83 &   & 58496.8180077(23) & 58412.8351305(14) &   & 58582.45486735(75) \\ 
	84 &   & 58582.352511(19) & 58582.1782558(12) &   & 58675.8824303(18) \\ 
	85 &   & 58585.4992656(36) & 58933.19292746(61) &   & 58753.9098546(79) \\ 
	86 &   & 58586.357471(23) & 59020.94654577(18) &   & 58844.7707124(12) \\ 
	87 &   & 58668.17308301(75) & 59036.90176275(19) &   & 59020.84561611(69) \\ 
	88 &   & 58754.851860(46) & 59076.06457880(74) &   & 59037.27243050(41) \\ 
	89 &   & 58760.859300(72) & 59351.29256505(59) &   & 59075.77276022(81) \\ 
	90 &   & 58846.6798719(48) & 59565.6005516(11) &   & 59146.6133467(18) \\ 
	91 &   & 58937.3636098(19) & 59599.68674527(20) &   & 59246.71413578(97) \\ 
	92 &   & 59029.1916226(28) & 59644.28887715(32) &   & 59317.5547223(18) \\ 
	93 &   & 59264.3399930(18) & 59705.2088602(36) &   & 59351.43499999(36) \\ 
	94 &   & 59325.5586680(37) & 59947.43818330(32) &   & 59362.2150910(84) \\ 
	95 &   & 59325.5586680(24) & 60110.97913770(26) &   & 59565.4968589(40) \\ 
	96 &   & 59454.003456(10) &   &   & 59599.3771697(16) \\ 
	97 &   & 59604.76158787(98) &   &   & 59644.55092666(54) \\ 
	98 &   & 59640.5201577(16) &   &   & 59705.124824(25) \\ 
	99 &   & 59700.308487(29) &   &   & 59813.9523949(92) \\ 
	100 &   & 59893.976925(16) &   &   & 59947.4203255(39) \\ 
    
    \bottomrule\\
   
   
   
\end{longtable}

 \begin{deluxetable*}{lccc}
\tabletypesize{\footnotesize}
\tablecaption{Timing Parameters for the Ter5 RBs.\label{bintab1}}
\tablewidth{0pt}
\tablehead{\colhead{Parameter} & \colhead{Ter5P} & \colhead{Ter5ad} & \colhead{Ter5ar}}
\startdata
Pulsar Name \dotfill & PSR J1748$-$2446P & PSR J1748$-$2446ad & PSR J1748$-$2446ar \\
\cutinhead{Data Reduction}
Span of Timing Data (MJD) \dotfill & 53204$-$60111 & 53320$-$59948 & 53193$-$59948 \\
Number of TOAs \dotfill & 1536 & 388 & 495 \\
RMS TOA Residual ($\mu$s) \dotfill & 40.7 & 31.4 & 66.5 \\
Reduced $\chi^2$ \dotfill & 1.01 & 1.03 & 1.02 \\
EFAC for incoherent data \dotfill & 2.52 & 2.65 & 3.47 \\
EFAC for coherent data \dotfill & 5.79 & 4.07 & 4.86 \\
\cutinhead{Timing Parameters}
Right Ascension (RA, J2000) \dotfill & $17^{\rm h}\;48^{\rm m}\;05\fs03815(7)$ & $17^{\rm h}\;48^{\rm m}\;03\fs8479(1)$ & $17^{\rm h}\;48^{\rm m}\;04\fs6196(2)$ \\
Declination     (DEC, J2000) \dotfill & $-24\degree\;46\amin\;41\farcs29(3)$ & $-24\degree\;46\amin\;41\farcs84(5)$ & $-24\degree\;46\amin\;45\farcs87(8)$ \\
Proper Motion in RA (mas\,yr$^{-1}$) \dotfill & $-1.8(1)$ & $-2.2(2)$ & $0.3(3)$ \\
Pulsar Spin Period (ms) \dotfill & $1.72861982757976(9)$ & $1.3959548139683(1)$ & $1.9528106824465(3)$ \\
Pulsar Spin Frequency (Hz) \dotfill & $578.49619913252(3)$ & $716.35556537627(6)$ & $512.08240972300(8)$ \\
Spin Frequency Derivative (Hz\, s$^{-1}$) \dotfill & $-8.66271(8) \times 10^{-14}$ & $1.74117(6) \times 10^{-14}$ & $6.76849(7) \times 10^{-14}$ \\
Frequency 2nd Derivative (Hz\, s$^{-2}$) \dotfill & $4(2) \times 10^{-26}$ & $-4(2) \times 10^{-27}$ & $-1.82(3) \times 10^{-25}$ \\
Frequency 3rd Derivative (Hz\, s$^{-3}$) \dotfill & $-2.5(2) \times 10^{-33}$ & $-$ & $-$ \\
Frequency 4th Derivative (Hz\, s$^{-4}$) \dotfill & $-4.3(1) \times 10^{-41}$ & $-$ & $-$ \\
Reference Epoch (PEPOCH, MJD) \dotfill & 56657.660271337430459 & 56633.74460856008227 & 56570.518685778268264 \\
Dispersion Measure (DM, pc\,cm$^{-3}$) \dotfill & $238.71(1)$ & $235.63(2)$ & $238.66(1)$ \\
DM Derivative (pc\,cm$^{-3}$\,yr$^{-1}$) \dotfill & $-0.007(1)$ & $-0.009(2)$ & $-0.001(1)$ \\
\cutinhead{Orbital Parameters}
Orbital Period (days) \dotfill & $0.362618545(8)$ & $1.09442881(5)$ & $0.513338066(9)$ \\
Orbital Period Derivative \dotfill & $1.38(8) \times 10^{-10}$ & $0.0(0)$ & $0.0(0)$ \\
Projected Semi-Major Axis (lt-s)  \dotfill & $1.271836(1)$ & $1.102814(3)$ & $1.498546(4)$ \\
Ref.~Epoch of Periastron ($T0$, MJD) \dotfill & $53800.32842747(6)$ & $53318.995701960084(6)$ & $53495.2744166(2)$ \\
\cutinhead{Derived Parameters}
Mass Function (\msun) \dotfill & $0.01679872(5)$ & $0.001202305(9)$ & $0.0137115(1)$ \\
Min Companion Mass (\msun) \dotfill & $\geq$\,0.38 & $\geq$\,0.14 & $\geq$\,0.35 \\
\enddata
\tablecomments{Numbers in parentheses represent 1-$\sigma$ uncertainties in the last digit
as determined by {\tt TEMPO}, {\tt PINT}, or via standard error propagation.
The timing solutions used the DE440 Solar System Ephemeris and times are all in 
Barycentric Dynamical Time (TDB), referenced to TT(BIPM2021). The eccentricity and longitude
of periastron, $\omega$, for each of the binaries were each assumed to be zero. Minimum 
companion masses were calculated assuming a pulsar mass of 1.4\,\msun.}
\end{deluxetable*}

 \begin{deluxetable*}{lcc}
\tabletypesize{\footnotesize}
\tablecaption{Timing Parameters for M28H and NGC 6440D.\label{bintab2}}
\tablewidth{0pt}
\tablehead{\colhead{Parameter} & \colhead{M28H} & \colhead{NGC6440D}}
\startdata
Pulsar Name \dotfill & PSR J1824$-$2452H & PSR J1748$-$2021D \\
\cutinhead{Data Reduction}
Span of Timing Data (MJD) \dotfill & 53629$-$60084 & 53478$-$59894 \\
Number of TOAs \dotfill & 866 & 574 \\
RMS TOA Residual ($\mu$s) \dotfill & 46.1 & 31.9 \\
Reduced $\chi^2$ \dotfill & 1.01 & 1.02 \\
EFAC for incoherent data \dotfill & 2.75 & 1.47 \\
EFAC for coherent data \dotfill & 7.37 & 1.32 \\
\cutinhead{Timing Parameters}
Right Ascension (RA, J2000) \dotfill & $18^{\rm h}\;24^{\rm m}\;31\fs6098(2)$ & $17^{\rm h}\;48^{\rm m}\;51\fs6457(1)$ \\
Declination     (DEC, J2000) \dotfill & $-24\degree\;52\amin\;17\farcs15(4)$ & $-20\degree\;21\amin\;07\farcs41(2)$ \\
Proper Motion in RA (mas\,yr$^{-1}$) \dotfill & $-1.1(3)$ & $-1.4(2)$ \\
Pulsar Spin Period (ms) \dotfill & $4.6294137643019(6)$ & $13.4958205400413(8)$ \\
Pulsar Spin Frequency (Hz) \dotfill & $216.01007188235(3)$ & $74.097013740888(5)$ \\
Spin Frequency Derivative (Hz\, s$^{-1}$) \dotfill & $-3.6139(2) \times 10^{-15}$ & $-3.22033(1) \times 10^{-15}$ \\
Frequency 2nd Derivative (Hz\, s$^{-2}$) \dotfill & $-8(9) \times 10^{-28}$ & $5(2) \times 10^{-28}$ \\
Reference Epoch (PEPOCH, MJD) \dotfill & 56856.714649097440997 & 56686.128079588925175 \\
Dispersion Measure (DM, pc\,cm$^{-3}$) \dotfill & $121.38(2)$ & $224.999(6)$ \\
DM Derivative (pc\,cm$^{-3}$\,yr$^{-1}$) \dotfill & $0.014(3)$ & $-0.004(1)$ \\
\cutinhead{Orbital Parameters}
Orbital Period (days) \dotfill & $0.43502746(1)$ & $0.2860686141(6)$ \\
Orbital Period Derivative \dotfill & $0.0(0)$ & $-1.7(1) \times 10^{-11}$ \\
Projected Semi-Major Axis (lt-s)  \dotfill & $0.719473(4)$ & $0.397212(1)$ \\
Ref.~Epoch of Periastron ($T0$, MJD) \dotfill & $53755.2263970(4)$ & $56479.176207(1)$ \\
\cutinhead{Derived Parameters}
Mass Function (\msun) \dotfill & $0.00211297(3)$ & $0.000822259(9)$ \\
Min Companion Mass (\msun) \dotfill & $\geq$\,0.17 & $\geq$\,0.12 \\
\enddata
\tablecomments{Numbers in parentheses represent 1-$\sigma$ uncertainties in the last digit
as determined by {\tt TEMPO}, {\tt PINT}, or via standard error propagation.
The timing solutions used the DE440 Solar System Ephemeris and times are all in 
Barycentric Dynamical Time (TDB), referenced to TT(BIPM2021). The eccentricity and longitude
of periastron, $\omega$, for each of the binaries were each assumed to be zero. Minimum 
companion masses were calculated assuming a pulsar mass of 1.4\,\msun.}
\end{deluxetable*}

\end{document}